# Reproducibility, Replicability, and Repeatability:
# A survey of reproducible research with a focus on high performance computing

Antunes Benjamin

LIMOS, CNRS, Clermont Auvergne INP Université Clermont Auvergne

Mines St-Etienne

benjamin.antunes@uca.fr

Hill David

LIMOS, CNRS, Clermont Auvergne INP Université Clermont Auvergne

Mines St-Etienne

david.hill@uca.fr

**ABSTRACT**

Reproducibility is widely acknowledged as a fundamental principle in scientific research. Currently, the scientific community grapples with numerous challenges associated with reproducibility, often referred to as the "reproducibility crisis." This crisis permeated numerous scientific disciplines. In this study, we examined the factors in scientific practices that might contribute to this lack of reproducibility. Significant focus is placed on the prevalent integration of computation in research, which can sometimes function as a black box in published papers. Our study primarily focuses on high-performance computing (HPC), which presents unique reproducibility challenges. This paper provides a comprehensive review of these concerns and potential solutions. Furthermore, we discuss the critical role of reproducible research in advancing science and identifying persisting issues within the field of HPC.

# 1 INTRODUCTION

Reproducibility is recognized as a cornerstone of science. In 1660, the Royal Society adopted the motto "*Nullius in verba*," which translates to "take nobody's word for it" (https://royalsociety.org/about-us/history/). Philosophers of science agree that reproducibility is one of the criteria that distinguishes science from pseudo-science. Since the 2010s, there has been a significant increase in the global interest among scientists in reproducible research. (Fanelli, 2018) shows an exponential increase in published papers on the theme of the reproducibility crisis. An increasing number of journals and conferences are concerned with the reproducibility of published articles (Drummond, 2018) (Bajpai et al., 2019). We also noted the creation of a journal dedicated to reproducing articles (Rougier et al., 2017). Reproducible research is a highly active and rapidly evolving field. We found one survey providing state-of-the-art reproducibility in scientific computing (Ivie and Thain, 2018), and several books attempting to do so (Desquilbet et al., 2019) (National Academies of Sciences, 2019) (Randall and Welser, 2018). Without minimizing the quality of previous works, we believe that, as this theme is evolving, it is pertinent to actualize our knowledge and provide an up-to-date state-of-the-art for the definitions and technologies used in reproducible research. We want to provide another perspective by focusing on high performance computing (HPC). Though previously cited works emphasize computational reproducibility and are more focused on the global scientific method or workflows like (Ivie and Thain, 2018), with HPC, we are on the frontline where this type of emerging problem has more impact.

Considering the increasing importance of reproducibility in scientific research, this study seeks to address the critical question of how reproducible research can be effectively achieved within the realm of high-performance computing (HPC). By examining the definitions and nuances of reproducible research, we delved into the multifaceted reproducibility crisis that spans various scientific fields, highlighting the specific challenges and opportunities presented by HPC. We analyzed the factors contributing to the loss of reproducibility, from open science and documentation to software engineering and workflow complexity. Moreover, we discuss the reproducibility problems unique to HPC, including issues with parallel computing, random number generation in parallel Monte Carlo simulations, optimization, hardware heterogeneity, and the emerging fields of quantum computing and machine learning. The study then explores a range of solutions designed to enhance reproducibility, such as versioning, literate programming, and advanced workflow management, while tackling HPC-specific challenges, such as floating-point reproducibility and error management. Through this comprehensive survey, we aim to actualize our understanding of reproducible research in HPC and contribute to the ongoing discourse on maintaining the integrity of scientific computation in an era in which the reproducibility of research is paramount.

First, we present the importance of reproducible research and the definitions of various terms used in this domain. We then demonstrate how the reproducibility crisis is occurring in several scientific fields. We also discuss the movement against the rise of reproducible research. Next, the reasons for the loss of reproducibility with a specific focus on high performance computing (HPC) are discussed. Finally, we present the current solutions to these different problems. Before concluding, we discuss the open problems in the area of reproducible research for high-performance computing applications.

# 2 DEFINITIONS OF REPRODUCIBLE RESEARCH

## 2.1 Evolution in terminologies

Three principal terms are employed in the field of reproducible research: reproducibility, replicability, and repeatability. Despite reproducibility being seemingly straightforward, a consensus definition among researchers and research fields has only been reached recently. Before 2020 (old definitions), the Association for Computing Machinery (ACM) defined these terms as follows:
- Repeatability: Same team, same experimental setup
- Reproducibility: Different teams, different experimental setups
- Replicability: Different teams, same experimental setup.

In these definitions, as Drummond stated (Drummond, 2009), "*reproducibility requires changes, replicability avoids it.*" Reproducibility implies that a different team applying a different method or setup for the same scientific question obtains the same scientific conclusions, thereby reinforcing the discovery. In contrast, replicability aims for a different team to achieve the same results with the stated precision using artifacts of the first team. In literature, authors were sometimes using the word "reproducibility" to refer to "replicability", and vice versa. Advised by (National Information Standards Organization), ACM changed its definitions after 2020 by swapping terms between reproducibility and replicability. The main reason for this is the need for a better match with the practices of other research fields. The new ACM definitions are equivalent to those proposed by the National academies of sciences, engineering and medicine, that defined: "*Reproducibility is obtaining consistent results using the same input data; computational steps, methods, and code; and conditions of analysis*" and "*Replicability is obtaining consistent results across studies aimed at answering the same scientific question, each of which has obtained its own data*" (National Academies of Sciences, 2019). To illustrate the problem of non-consensus on definitions, we examined the literature. Herein, we regard the definition D1 as that employed by the ACM before 2020 and the definition D2 as that adopted by the ACM after 2020.

In an article discussing reproducibility in computer science (Hinsen, 2014), Hinsen adopted the definition D1. For the article describing the creation of their journal ReScience, (Rougier et al., 2017) used the definition D2. In a presentation in 2017 (Hinsen, 2017), Hinsen modified his 2014 definition to include D2. In their paper, (Stanisic et al., 2015) use the definition D1. An article (Cohen-Boulakia et al., 2017) on the theme of life sciences is in category D1. (Drummond, 2009) used the definition D1. In the study of reproducibility, (Collberg and Proebsting, 2016) applied the definition D1. (Gundersen and Kjensmo, 2018) used the definition D1. In the network domain, (Bajpai, Brunstrom, et al., 2019) used the definition D1. (National Academies of Sciences, 2019) is based on the D2 definition and has a strong impact (more than 600 citations to date). A more recent reproducibility survey in 2018 (Ivie and Thain, 2018) also used the definition D1. Stodden used the 2011 (Stodden, 2011) D1 definition and then switched to D2 in 2014 (Stodden et al., 2014). Several articles still define the terms reproducibility, repeatability, and replicability." Some of them even uses the notion of "reproducible research" without explicitly bringing nuances to the different notions.(Hill et al., 2013)(Hill, 2015)(Hill et al., 2017)

In our small sample, which predominantly comprises highly cited papers and authors active in the field, we identified eight articles that employed the definitions provided by the ACM prior to 2020 (D1); they were published between 2009 and 2019. Four studies used the new ACM definitions (D2). From our perspective as computer scientists, it appears that the majority of published work relies on outdated ACM definitions. A noteworthy study by (Barba, 2018) found that the scientific field used different definitions. From this study, we realized that Computer Science is a one of the only field to use the definition given at the beginning of this section.

Nevertheless, to accurately define a concept, it is imperative to consider perspectives beyond its immediate domain (computer science). Some definitions of reproducible research are specific to each domain, which we want to avoid in the context of reproducibility, as we want to standardize as much as possible between the disciplines. However, the trend towards adopting newer definitions is discernible among authors actively contributing to the reproducible research domain.

The fact that computer science was one of the only fields using these definitions was an argument to follow the NISO standardization advice; the ACM switched its reproducibility and replicability definitions. The 2020 and current definitions of both terms are as follows:

"*Reproducibility (Different team, same experimental setup): The measurement can be obtained with stated precision by a different team using the same measurement procedure, the same measuring system, under the same operating conditions, in the same or a different location on multiple trials. For computational experiments, this means that an independent group can obtain the same result using the author's own artifacts.*

*Replicability (Different team, different experimental setup): The measurement can be obtained with stated precision by a different team, a different measuring system, in a different location on multiple trials. For computational experiments, this means that an independent group can obtain the same result using artifacts which they develop completely independently.*

*Repeatability (Same team, same experimental setup): The measurement can be obtained with stated precision by the same team using the same measurement procedure, the same measuring system, under the same operating conditions, in the same location on multiple trials. For computational experiments, this means that a researcher can reliably repeat his own computation.*" ([ACM badges](#)).

We observe that some authors such as Stodden or Hinsen, active in the reproducibility research field in computer science, were using previous definitions. They swapped their definitions of reproducibility and replicability, which now match the ACM 2020 update.

The last term to discuss is repeatability. This has led to less controversy than the other two approaches. However, in computer science research, there is sometimes certain confusion regarding repeatability and reproducibility. We discuss its importance later in this paper.

However, from a philosophical perspective, reproducibility is the only term that encompasses all the different notions. Karl Popper, a famous philosopher of science, discussed the logic behind scientific discoveries at the beginning of the 20$^{th}$ century, first in German (1934/35) and then in English in 1959. A recent re-edition is available from (Popper, 2005). Popper defined reproducibility as a criterion for distinguishing between science and pseudoscience (Hill, 2019). The use of other terms is still useful as we want to describe them more precisely. More focused on the reproducibility term, (Goodman et al., 2016) proposed three definitions that could fit and encompass the standard definition:

- Method reproducibility: refers to the capacity to faithfully replicate experimental and computational procedures by employing identical data and tools to achieve consistent results. This aligned with the revised ACM definitions of reproducibility and repeatability.
- Results reproducibility: pertains to the generation of consistent findings in a novel investigation using identical experimental methods. This corresponds to a new definition of ACM reproducibility. However, to define reproducibility, we used exactly the same materials (artifacts) to generate the same results.
- Inferential reproducibility: involves reaching qualitatively similar conclusions, either through the independent replication of a study or by reanalyzing the results of the original study. This corresponds to the new definition of ACM replicability.

With these definitions, we maintain the main term reproducibility, but add the context of its usage. However, we consider that the ACM post 2020 definition is now the standard that authors should adhere to as an objective of standardization.

## 2.2 Importance of reproducible research

One of the origins of the reproducible research movement in computational science was the work of Claerbout in 1992 (Claerbout and Karrenbach, 1992). (Buckheit and Donoho, 1995) synthetized it with the famous citation: "*An article about computational science in a scientific publication is not the scholarship itself, it is merely advertising of the scholarship. The actual scholarship is the complete software development environment and the complete set of instructions which generated the figures*". (Marwick, 2015) highlighted the challenge posed by computer programs: they act as black boxes. Currently, almost all scientific research involves the use of computers. In the not-so-distant past, it was necessary to perform all the calculations, all the transformations, and use of data, and so on, and to precisely describe the procedure in a research article to be published. This made it feasible to reproduce the articles in a simple manner. Currently, complex software layers hide numerous elements. It is not easy to know the purpose of these layers and elements and it becomes much more complicated to reproduce the results published by others if the same machine is not available with the same software stack. The computer is becoming a research instrument on its own. Therefore, it should undergo the same quality checking (meticulous work of metrology) that is employed by biologists or physicists on their instruments.

As Popper stated, reproducibility is mandatory for scientific advancement. We should be able to use the artifacts of this study to repeat the experiment and obtain the same results. We need research paper artifacts to avoid mistakes or fraud, much less frequently. Reproducibility involves an independent research team conducting an experiment based solely on the documentation provided by the original research team. The ability to facilitate the reproduction of the same results will increase trust in the published results. Notably, other researchers can be expected to maintain a higher level of objectivity as they have no interest in exaggerating the performance of a method developed by other scholars. In addition, these researchers may not share the same preconceptions and tacit knowledge as the initial team that reported the research. In addition, variations in hardware and software configurations among different researchers further aid in controlling noise variables associated with hardware and ancillary software, as well as implicit knowledge and preconceptions. However, this last point is true only when considering that reproducibility meets with the stated precision and not with bitwise identical results, which might be the objective for some authors and also a requirement for debugging.

Regarding repeatability, definitions can vary across scientific fields. In fact, in computer science, machines are designed to be deterministic (except for quantum computing machines or simulators). Using digital computing, identical bitwise run-to-run results are to be obtained when we are using the same machine for the same program. This point has been assumed by many scientists; however, this is not always the case, particularly when dealing with high-performance computing. However, this is essential for debugging and trusting the use of deterministic computers. Repeatability is a significant concern for researchers who are aware of debugging, and this activity can be particularly difficult with parallel computing. Ensuring reliable parallel debugging requires repeatability and identical bitwise results. And in this sense, the ACM definition cannot be fully agreed with, as it adds that results are identical with a "stated precision." This is because the ACM definition of repeatability is derived from the International Vocabulary of Metrology. In our opinion, this definition is perfectly correct for quantum computer science, but not for classical deterministic computing where "bitwise identical results" are needed to debug properly. In classical deterministic computer science, the "stated precision" should allow no difference. Nevertheless, the ACM definition remains valid if "perfect" precision with identical results is required.

Finally, replicability is a mandatory scientific requirement. Indeed, the more a scientific hypothesis is replicated worldwide (with different research teams and different methods or experiments), the stronger the hypothesis becomes and the more it will be shared among all scientists. This is the heart of science and confidence in scientific conclusions.

We can observe three terms in the ACM definitions standing at different levels. Repeatability stands at the author's level, who must debug or redo his/her own experiments. Reproducibility stands at the paper level: other researchers might want to rely on this paper to build their own research, and pursue the goal of improving knowledge avoiding the "reinvention of the wheel." Achieving this goal implies that papers are published with all their artifacts, which improves confidence in the published results. Finally, replicability stands at the science level: different research teams perform different experiments but obtain the same scientific conclusion. This is necessary to validate scientific hypotheses.

All these notions are crucial for the scientific community. There has been an increase in the interest in reproducible research, which can now be observed in scientific conferences and journals. The creation of a journal dedicated to reproducing the results of published papers is commendable in this regard (Rougier et al., 2017). In his article (Drummond, 2018), Drummond, even if he does not approve it, assessed the fact that many conferences (AAAS, AMP, ENAR, NSF, SIAM-CSE, SIAM-Geo) and journals, in the field of machine learning, are having increased concern regarding reproducible research. (Stodden et al., 2018) evaluated the effectiveness of a journal policy requiring authors to make data and codes available upon request post-publication to promote reproducibility. From a random sample of 204 scientific papers published in high-impact science journals after implementing this policy, artifacts were obtained from only 44% of the sample. These findings were successfully reproduced in 26% of cases. This policy is certainly an improvement over having no policy, but it is still insufficient to ensure reproducibility. Stodden et al. are assessing whether conferences and journal policies improve reproducibility and are continuing their research in this direction.

## 3  REPRODUCIBILITY CRISIS: EXAMPLES FROM DIFFERENT DOMAINS

The crisis of scientific reproducibility is now a global and widely transdisciplinary phenomenon that contributes to society's distrust of the world of research. In 2016, Baker (Baker, 2016) published a survey of 1576 scientists to determine their opinions on reproducibility crisis. Approximately 90% of respondents were of the opinion that a significant or slight reproducibility crisis existed. Only 3% were convinced that there was no crisis. This study emphasizes the consensus among the scientific community that a reproducibility crisis spans various disciplines.

The existing literature offers a myriad of theoretical and empirical examples that highlight the reproducibility crisis. In medicine, we have a famous provocation from Ioannidis, a top epidemiologist. One of his major articles in this domain was entitled "Why most published research findings are false?" (Ioannidis, 2005). In this highly-cited paper, he discussed the statistical flaws that might affect the published results. (Errington et al., 2021) presented reproducibility results for cancer studies, showing a success rate of only 46% out of 112 attempts. (Begley and Ellis, 2012) raised a reproducibility concern for cancer research. This rate may seem relatively optimistic considering (Ioannidis, 2015), who claimed that 85% of research funding was being wasted. (Eklund et al., 2016) identified a significant problem with MRI studies, indicating that several articles may have reported false results. (Open Science Collaboration, 2015) revealed a similar problem with psychology articles, with reproducibility rates ranging from 30 to 50%. In the field of neuroscience, (Topalidou et al., 2015) were unable to reproduce a model and had to spend three months reimplementing it.

This crisis has heavily impacted the medical field in areas such as medication, medical devices (MRI), psychology, cancer research, and neuroscience. However, as aforementioned, no field is an exemption. Computer Science is supposed to be an exact science, dealing with deterministic machines and hence, not expected to face this issue. Nevertheless, computer science is not immune to a reproducibility crisis; in fact, it is partly an instigator. A thorough study by (Collberg and Proebsting, 2016) on the reproducibility of computer science articles yielded insufficient results, with only approximately 30% of the 601 research papers examined being reproducible. (Manninen et al., 2017) attempted to reproduce four models of "calcium excitability in astrocytes," and 3 out of 4 models lacked essential information, and 2 out of 4 models had incorrect equations. Even after correcting the different models, they did not produce consistent results. (Mesnard and Barba, 2017) studied fluid mechanics and found that it took three years to reproduce the results obtained from their own tools using two other tools and a parallel version of their tool. As demonstrated by (Gundersen and Kjensmo, 2018), artificial intelligence is also not an exemption. Further, a recent study by David et al. demonstrated that machine learning is undecidable (Ben-David et al., 2019). In the networking domain, (Kurkowski et al., 2005) found that less than 15% of papers on MANET (Mobile ad hoc networks) network simulations were reproducible. In image processing, (Kovacevic, 2007) studied 15 published papers in her field and found that none of the presented algorithms were supported by any code, and only 33% of the data were available. (Vandewalle et al., 2009) examined 134 papers in the same field and found that 9% of the papers had codes available, and 33% had data.

This highlights that computer science, across its various subdomains, is not immune to reproducibility challenges. These findings highlighted the importance of developing reproducible research methods. Emphasizing the need for reproducible research can prevent fraud and scandals. For instance, Reinhart and Rogoff, world economy specialists, claimed in 2010 that increasing a country's debt by over 90% of its Gross Domestic Product would stop its economic growth. This assertion led several occidental countries, such as the United States, to adopt austerity policies. However, (Herndon et al., 2014) later proved that this was incorrect. Reinhart and Rogoff excluded data that contradicted their findings, which resulted in calculation errors. Because they shared codes and data, it was possible to check their work. This strongly advocates reproducible research and open science (sharing all data related to an article). The only way to demonstrate that an article is false is by accessing its artifacts. Another example described by (Miller, 2006) involves Geoffrey Chang, who had made significant contributions toward antibiotic-resistant bacterial protein structure. However, several years later, his results were found to be incorrect owing to the discovery of programming errors in internal tools used by Chang.

Recently, the Covid19 crisis has greatly increased the awareness of non-scientist citizens regarding the importance of reproducible research. We have witnessed scandals in the context of Covid-19, such as the retraction of two papers from the Lancet and New England Journal of Medicine (Piller and Servick, s. d.). Both studies influenced international policies regarding the use of certain drugs, and they had to be quickly retracted. Another case was Neil Ferguson's COVID model (Ferguson et al., 2020), for which (Pouzat, 2022) prepared a humorous article highlighting the scandal caused by the nonpublication of Ferguson's initial code. This model influenced international policies adopted on lockdown measures, particularly in England. With international pressure, mainly from the US, Neil Fergusson published a revised version of his code, which was then reported as severe flaw (https://www.telegraph.co.uk/technology/2020/05/16/coding-led-lockdown-totally-unreliable-buggy-mess-say-experts/), leading to a petition on GitHub (https://github.com/mrc-ide/covid-sim/issues/165) to remove the code to avoid its use as a basis for other epidemiological models. Financial pressure, particularly in medicine on a world scale (Abbasi, 2020), can significantly decrease the quality of science, as without reproducibility, the endeavor treads closely like pseudoscience. Funding and conflicts of interest gangrene these situations, despite the need for fast and worldwide international public collaboration. (Iqbal et al., 2016) stated that "*Articles published in journals in the clinical medicine category versus other fields were almost twice as likely to not include any information on funding and to have private funding*". Moreover, it is known that papers with industrial funding or industrial authors are less likely to share codes and data (Collberg et al., 2015), which lead to the fact that we cannot completely rely on industrial papers, if artifacts cannot be accessed.

## 4   OPPOSITION TO THE REPRODUCIBLE RESEARCH MOVEMENT

Not all authors agree with this reproducible research trend. In his article (Drummond, 2018), Drummond asserted that sharing the source code of an article is unnecessary. He believes that researchers are forced to do so to avoid getting a bad label but that it does not serve science. For him, the reproducible research movement was not based on facts, but only on intuition. He adds that the obligation to provide the source code will lead to papers being accepted based on technically weak criteria and that, according to his opinion, fraud has always existed and never posed a significant problem. However, Drummond supports the concept of open science. We disagree with Drummond's statement that sharing code and data has now become straightforward with the plethora of tools available to us. Why do we trust published articles automatically? We should be able to verify that what has been published is free of errors. The case previously discussed showed how an Excel error (at minimum) in an invited research paper published by trusted top scientists impacted economic policies of several countries (Herndon et al., 2014). It is rare to observe fraud. However, who does not make mistakes? There is the potential for authors to use the data selectively or make inadvertent errors. An increased number of reviews can enhance the detection of these errors. A bug-free code is a code that has not yet been sufficiently tested. Finally, regarding the claim that frauds have not posed significant problems, we think that it is up for debate. Furthermore, concerning publicly funded research, presumably financed by taxpayers, there seems to be an ethical imperative to ensure the full accessibility of outcomes. In a second study, (Drummond, 2019) strongly criticized the prioritization of article replication over novelty. This contradicts prevailing sentiments. Therefore, he opposes the changing customs of journals and conferences that are currently underway. There is also (Fanelli, 2018) that claims that the reproducibility crisis is widely exaggerated or even false and that this narrative is harmful as it demotivates young researchers. However, being rigorous is part of our job, and it should not deter young researchers.

Many researchers strongly disagree with contrarian voices, though useful for questioning the relevance of such an approach, as cited above (Drummond and Fanelli). (Stodden et al., 2013), (Bajpai et al., 2017), (Pouzat, 2022) in his humorous dialogue, (Rougier et al., 2017) in the creation of their journal dedicated to article replication, (National Academies of Sciences, 2019), (Ten Hagen, 2016), and others defend the idea that journals should encourage reproducible research. (Barba, 2018) stated regarding the definitions of reproducibility and replication, after Drummond's 2009 article: "*They, in turn, based their definitions on the emphatic but essentially flawed work of Drummond (2009)*." We believe that skepticism is a hallmark of competent scientists. Promoting a culture of inquiry and skepticism is crucial, yet it is vital to exercise caution, as the misuse of the "scientific" doubt has also been employed to impede the recognition of groundbreaking discoveries, as it was with the case of the link between cigarettes and lung cancer.

## 5   WHY DO WE LOSE REPRODUCIBILITY?

In our exploration of the challenges to reproducibility in computational research, we encountered a multifaceted landscape, as depicted in Figure 1. Loss of reproducibility can stem from various factors that are categorized as scientific computing and high-performance computing, each influenced by global context and user dependencies. In scientific computing, reproducibility issues arise primarily because of inconsistencies in software environments, workflows, prevailing scientific culture, and the degree of openness in science. These factors are complemented by the robustness of software engineering practices and quality of documentation and statistics. However, high-performance computing faces a unique set of challenges, including silent errors, the use of pseudo-random number generators (PRNGs), optimization techniques, parallel computing intricacies, and inherent complexities of quantum computing. These elements collectively highlight the intricate and layered nature of reproducibility loss, which necessitates a thorough understanding and strategic mitigation approach.

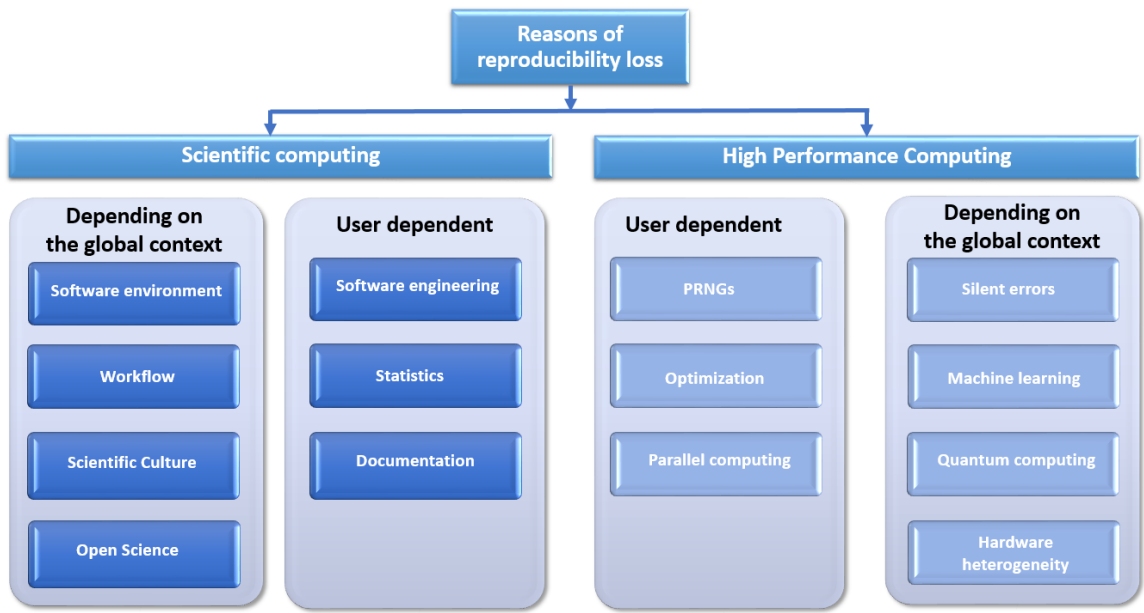

Figure 1: Overview of causes of reproducibility loss.

## 5.1 Open Science

A primary reason for the inability to reproduce articles is the reluctance or failure of the authors to share their artifacts. Without sharing code and data, the reproducibility of scientific work is impossible. The rise of computer science in all scientific fields has contributed to the reproducibility crisis, as described by Marwick (Marwick, 2015). Most authors advocating for reproducible research emphasize the importance of sharing both codes and data. Numerous studies cited earlier indicate that reproducibility issues predominantly arise from the absence of shared codes and data.

However, the use of proprietary software can lead to reproducibility problems. (Nüst et al., 2020) recalled that we need an open source for reproducible science. The National Academy of Sciences, Engineering and Medicine (National Academies of Sciences, 2019) stated that funding agencies should finance open sciences. The second aspect is the importance of the intellectual property of the published code. Without explicit licensing, default intellectual property rights apply, as outlined by (Halchenko and Hanke, 2015).

Finally, the authors informally and personally mentioned other reasons (Collberg and Proebsting, 2016). For example, some authors are hesitant to publish their code because they believe that it should be cleaned or modified. They may consider their code insufficiently clean for sharing, which is unfortunate. At times, the code was not initially intended for sharing, and making it easily accessible would require too much effort on the part of the author, who may not always be up-to-date with the best practices in software engineering (SE). Licensing issues can prevent authors from publishing their code, particularly when dealing with non-open-source software. Moreover, a difference in code accessibility was observed between the public and private domains. Some authors may no longer use their own code because the only person who knew how to use it may have left the team. This highlights the importance of implementing effective SE practices to avoid these pitfalls. In addition, the code can be lost more conventionally. This may be due to crashes or bad backup policies;

sometimes, the programming language or style may have become too obscure or outdated for use. In rare cases, non-sharing is intentional to mitigate risks, such as disclosing a security vulnerability and to prevent exploitation by others.

The main way to promote the open science movement was to discuss it. Several studies on reproducible research have highlighted the importance of open science. However, there are concrete examples of actions taken to enhance the opening-up of science. One of the first open science projects was Jon Claerbout's Stanford Exploration Project. More recently, CERN developed Zenodo in the framework of open science. The journals also reacted, as shown by PLOS Politics (https://journals.plos.org/plosone/s/materials-and-software-sharing). States, as they have financing power, must involve themselves in open science, such as the National Institute of Health (Collins and Tabak, 2014) and the French government (https://www.ouvrirlascience.fr/second-national-plan-for-open-science/ ). Finally, integrating open science and creating awareness on reproducibility in an educational setting has the potential to further this cause (Janz, 2016).

### 5.2 Documentation

Documentation in the context of scientific research refers to the process of recording and providing detailed information on the code, software, methods, and experimental procedures used in a study.

In the context of reproducibility in computer science research, the importance of documentation while sharing code cannot be overstated. Precise and comprehensive documentation is essential to facilitate the replication of research findings and promote transparency and credibility in the scientific community. As highlighted by (Boettiger, 2015), incomplete or imprecise documentation on how to install and run code can be a significant barrier to replication, particularly for researchers who may not be familiar with the specific tools and package managers involved. This problem is further emphasized by (Kitzes et al., 2018) in their work, where the availability of data and software, including the importance of proper documentation, open-source practices, software engineering techniques, and copyright considerations are discussed.

Best practices, as outlined by (Wilson et al., 2014), emphasize the need to document the design and purpose of a code, rather than focusing solely on its mechanics. By documenting the interfaces and reasons, researchers can enable others to understand the functionality of a code and its objectives, thereby facilitating reproducibility. The study by (Boettiger, 2015) also points out the impact of imprecise documentation on reproducing analyses, with a significant number of experiments using popular software being irreproducible owing to incomplete parameter documentation.

Moreover, in the rapidly evolving field of artificial intelligence (AI), documentation practices are critical to ensure reproducibility, as highlighted by (Gundersen and Kjensmo, 2018). Several AI research studies lack well-documented methods and codes, making it challenging to reproduce the reported results and hindering progress in the field. However, this study also acknowledges that documentation practices have improved over time, underlining the importance of continued efforts to promote and maintain rigorous documentation standards.

The effective documentation of software and code is a fundamental aspect of reproducible research in computer science. Researchers should strive to provide clear and comprehensive documentation, including installation procedures, parameter descriptions, and design rationales. Proper documentation not only enables the reproduction of research findings but also fosters collaboration, knowledge sharing, and advancement in science.

### 5.3 Statistics

In 2005, Ioannidis's (Ioannidis, 2005), barely hit the world of reproducible research. He stated that simulations demonstrate that research claims are more likely to be false than true in many study designs and settings, raising questions about the accuracy of the claimed research findings and their implications for research conduct and interpretation. The importance of statistics in reproducible research is undeniable. Statistics plays a crucial role in detecting and avoiding false-positive findings, which is a significant concern in scientific research. Several problems in statistics can lead to a loss of reproducibility in articles and replicability (the ability to obtain the same scientific conclusion with another experiment).

P-hacking, which is defined as the manipulation of statistical analyses to achieve significant results, is one of the main factors contributing to the reproducibility crisis. P-hacking practices such as selectively choosing data or conducting multiple tests until a significant result is obtained can lead to biased and unreliable conclusions. As described by (Forstmeier et al., 2017), this problem is exacerbated by cultural pressure to publish only significant findings. They showed that decreasing the sample size, increasing the pursuit of novelty, and engaging in multiple testing can all increase the probability of false-positive conclusions. Further, "*incorrect P-values due to unaccounted pseudoreplication, i.e., non-independence of data points*", can also contribute to this issue (Forstmeier et al., 2017). To address these issues, it is essential to adopt rigorous research practices. Pre-registering studies, blinding observers during data collection and analysis, and reporting all results regardless of their significance are strategies that can improve the objectivity of scientific research. Furthermore, shifting efforts from seeking novelty and discovery to reproducing important research findings could benefit the scientific community. It is crucial to evaluate research based on scientific rigor rather than relying solely on impact metrics. The reliance on p-values as a measure of evidence and significance is another critical aspect to consider in the context of reproducible research. P-values are often used to determine the strength of evidence in research findings; however, studies have questioned their reliability and objectivity (Nuzzo, 2014). Researchers have noted that even minor changes in statistical significance can result in significant changes in the interpretation of the results (Gelman and Stern, 2006). This discrepancy can lead to misleading conclusions and emphasizes the need for a reevaluation of statistical philosophy and methodologies (Nuzzo, 2014).

Hypothesizing after the results are known is named as "HARKing;" it is another common practice in scientific communication that can undermine reproducibility (Kerr, 1998). HARKing involves presenting post hoc hypotheses as if they were a priori hypotheses in research reports. This practice can lead to biased and inaccurate interpretations of data as it allows researchers to selectively choose hypotheses that align with their results. Although the motivations behind HARKing vary, they are widely considered inappropriate and have negative implications on scientific integrity.

It is also possible to overinterpret statistical results that are significant. In (Gelman and Stern, 2006), authors highlighted that even major changes in significance levels may correspond to minor, non-significant changes in the underlying quantities. This error is conceptually different from other issues related to statistical significance, such as practical importance, dichotomization into significant and non-significant results, and arbitrary threshold selection. The ubiquity of this statistical error calls for increased awareness among students and practitioners to avoid misinterpretation.

Some tools are developed to help with statistics, such as R Markdown used to simplify reproducible statistical analysis, making it suitable for both advanced research and introductory statistics courses (Baumer et al., 2014), or (Pernet et al., 2013), which introduces an open-source MATLAB toolbox designed for robust correlation analyses. The traditional Pearson's correlation, which is predominantly used in psychology research, is often limited to linear associations and is highly susceptible to outliers, which can distort data interpretation. The proposed open-source MATLAB toolbox offers alternative methods, namely percentage-bend correlation and skipped correlations, that counteract the effect of outliers either by downweighting or removal. These techniques yield better estimates of true associations and maintain accurate control over false positives without compromising statistical power.

## 5.4  Scientific culture

The current scientific publication landscape is plagued by several issues that impede the reproducibility of research, which are critical for the advancement of science. One prominent problem is the lack of incentives from journals and conferences to provide all artifacts related to a research article, such as datasets, codes, or detailed methods, which are essential for full transparency and reproducibility. Nosek wrote, "*Because of strong incentives for innovation and weak incentives for confirmation, direct replication is rarely practiced or published*," and "*Innovative findings produce rewards of publication, employment, and tenure; replicated findings produce a shrug*." (Open Science Collaboration, 2012), as quoted by (Collberg and Proebsting, 2016). Moreover, these venues often fail to encourage the reproduction and replication of studies, placing an overemphasis on novelty rather than robustness of the findings. This creates a culture that discourages reproductive efforts, often leaving early career researchers to withdraw from such endeavors in favor of activities that enhance their academic profiles. This issue is further exacerbated by the entrenched "Publish or Perish" mentality that puts pressure on researchers to continually produce new findings, sometimes at the expense of rigorous, quality research. The peer-review process is also not without flaws. Often, it does not prioritize the reproducibility of studies, which can lead to the publication of findings that cannot be verified independently. Finally, publication bias towards positive results often results in an underrepresentation of negative or null findings, further skewing the research landscape. Collectively, these issues pose significant challenges to the integrity and reliability of scientific research and call for systemic changes in the way science is conducted and communicated. (Bajpai et al., 2017) discussed three elements that pose problems for reproducibility: the lack of incentives from journals (which prioritize only innovative papers), a double-blind review process that requires authors to hide potentially crucial information or data, and reviewers who do not test for reproducibility. This article proposes including a reproducibility section to encourage authors, promote reproducible papers, and improve review processes. (Baker, 2016) presented 14 factors that can contribute to the loss of reproducibility, mainly related to statistical issues, publication pressure, and unavailable codes. It offers 11 solutions that directly address problems, such as improving statistical understanding. (Munafò et al., 2017) identifies cognitive biases, methodology improvements, increased collaboration among researchers, and enhanced peer review as problems and solutions to improve reproducibility. (Ten Hagen, 2016) argues that journals excessively favor novelty, which undermines reproducible research as it discourages researchers from attempting to replicate previously published results. (Fanelli, 2010) showed that funding and publication pressures push researchers to publish only positive results, limiting the publication of negative results that could also contribute to scientific progress by revealing what does not work or is not true, including negative results from replicating previously published articles.

## 5.5 Software environment

The software environment is a critical aspect of reproducible research as the use of computation has become ubiquitous in science. A large number of studies now use code, scripts, or data as inputs to generate outputs.

Even if you share your code and data, it is highly unlikely that another person attempting to reproduce the results will have the same software environment. Consequently, they may not be able to run the code properly because of potential incompatibility, different libraries, or unavailable software (Hinsen, 2013). To address this issue, Hinsen suggested using trusted and proven libraries, writing code with clarity (while being mindful of performance, particularly in the HPC context), documenting the formats used, evaluating dependencies, and providing ready-to-run examples to facilitate adoption. In 2014, Hinsen provided valuable insights and advised caution regarding the addition of unmentioned input data to workflows and potential software or hardware bugs (Hinsen, 2014). (Desquilbet et al., 2019) discussed these challenges in their work. The software environment layer comprises of several levels. First, an operating system is used. Computations can differ through the use of different operating systems, because the usage of hardware can differ. The use of open-source Linux-based systems is mandatory for reproducible research, as open-source software is required (unlike Windows or macOS, which are not open source, hence, it is difficult to obtain the same configuration to ensure reproducibility). Linux is strongly advised as it is widely used, fully open source, and offers several tools to conduct reproducible research. Second, as previously mentioned, codes always use libraries. Thus, when shared, the code does not work if the destination machine does not have access to the same libraries with the same version. Each library may also have dependencies with others. This is known as "Dependency Hell" (Boettiger, 2015). High-level tools, such as frameworks or libraries, may also act as black boxes that do not allow other researchers to use the code for their own research. This is what Hinsen called "reusable code VS re-editable code" (Hinsen, 2018). Compilers are important software in high-performance computing (HPC). Intensive computing requires avoiding the wastage of computing time and energy, and compiled languages are known to be more efficient than interpreted languages. The code executed on computing clusters or supercomputers is primarily produced using C, C++, or Fortran compilers. Numerous scientists who do not specialize in computer science, such as biologists, might find it easier to work with Python or R at their scale, and these are useful languages with many libraries. Different versions of compilers or programming languages can also lead to a loss of reproducibility. The importance of the entire software stack has already been clarified by Claerbout, according to the aforementioned Donoho's citation (Buckheit and Donoho, 1995).

### 5.6 Workflow

The workflow problem in reproducible research revolves around the need to capture and communicate the entire process of data analysis, experimentation, and code execution in a clear and organized manner. Most of the time, a paper does not use one code or script, but a succession of codes is applied to different data. A reproducible research workflow should enable other researchers to independently verify and reproduce the results of this study. However, in practice, workflows often lack transparency, making it difficult for others to understand the steps taken, parameters used, and data transformations applied. This lack of clarity can result in incomplete or ambiguous documentation, which makes it challenging to reproduce the exact sequence of operations that led to the reported findings. Furthermore, workflows can involve multiple tools, libraries, and software dependencies, and managing the interactions between these components adds complexity to the reproducibility of the process. The problem is compounded when workflows are spread across various scripts, notebooks, and different programming languages, as it becomes more difficult to ensure that every detail is captured accurately and consistently. In addition, version control is crucial for managing workflow changes over time. Without proper versioning, it may be difficult to trace back to the exact state of a workflow when a particular result was obtained.

Addressing the workflow problem in reproducible research requires adopting best practices in documenting the steps taken, providing clear explanations of the rationale behind the decisions, and ensuring the availability of all necessary data and codes. Cohen-Boulakia et al. conducted an exhaustive study on workflows, focusing on life sciences (Cohen-Boulakia et al., 2017). Workflows can take different forms as they are a large term. (Stanisic et al., 2015) also focused on this topic. We discuss these tools later in this paper. Figure 2 presents a visual representation of the workflow concept. This illustration shows a standard workflow sequence comprising stages, such as data gathering, data preparation, crafting software solutions, and conducting data analysis, culminating in the creation of a scientific paper. Replicating a study necessitates a scientist's capability to re-enact the entire sequence, referred to as the workflow. The precise outcome at each stage depends on the use of a particular software environment.

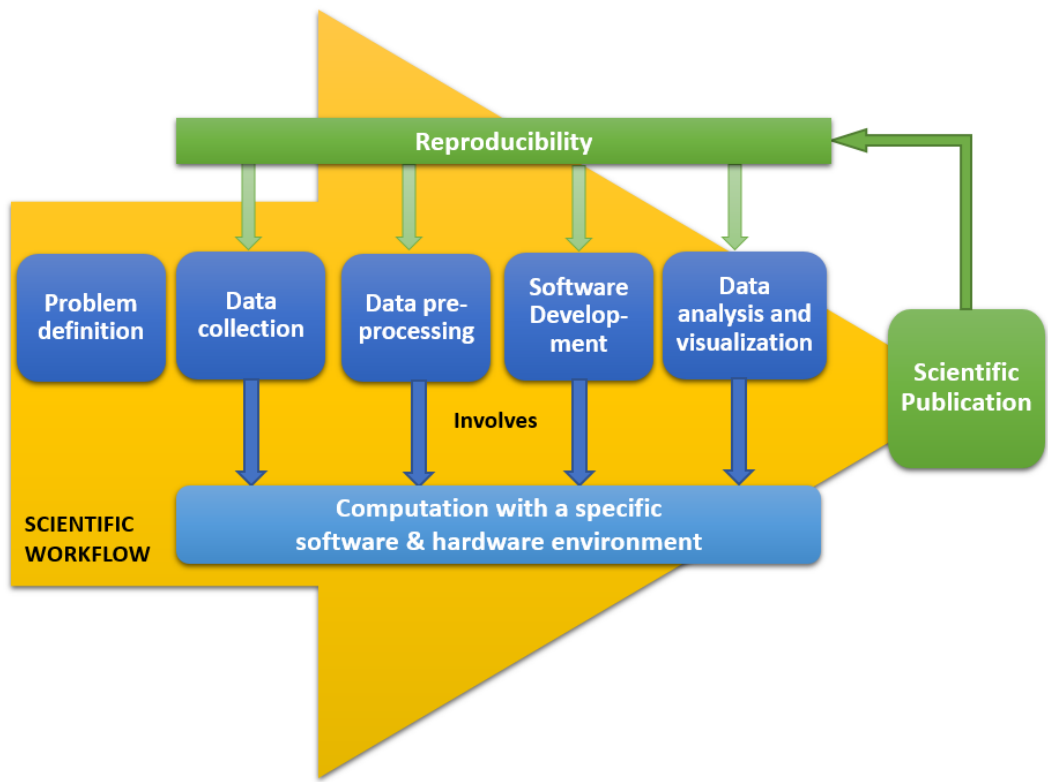

Figure 2: An example of scientific workflow, leading to a publication.

## 5.7 Software engineering

Software engineering has been frequently overlooked. As computer science has become ubiquitous, several scientific fields have used programming to help students conduct their research. It is now feasible to use code to prepare data, transform data, visualize data, make statistics, draw experiments, and so on. For high-performance computing, low-level skills are mostly needed (bash scripting, OS knowledge, and computationally efficient languages such as C, C++, or Fortran). For more general purposes and data analysis, scientists mainly use MATLAB, Python, R, high-level frameworks, libraries, and so on. All these skills can be difficult to acquire for those who follow other science streams subjects, for example, classical biology or physics. Although high-level technologies offer an easier way to handle computations, they are also less stable and more obscure than low-level technologies because they hide complexity in black boxes and harden reproducibility. These are "flaws" in the advantages of modern high-level technologies.

While publishing and providing artifacts is good, software engineering is mandatory to have a maintainable, readable, and evolvable code. Several tools have been developed to ensure reproducibility. However, these methods often require specific skills. Non-computer scientists use computers as a tool without sufficient knowledge regarding its use (assuming the absolute determinism of such a machine, for example). (Hinsen, 2018) presented an example of the importance of knowing what is happening behind the scenes. He read a dataset with health data from a CSV file using Pandas software (Data related to five years was considered). Pandas automatically attempts to find the appropriate data types for the data, the datatype for this data was "int64." When he loaded the full dataset (33 years vs 5 years), data were recognized as if they were now of the type "object" and not "int64" anymore. In Pandas documentation, it mentions "intelligent conversion of tabular data," but rules are not specified. Though this can be handled with some effort, this shows that there exists hidden complexities that nonexperts might not think of. The study also pointed out that reusable and re-editable codes are not the same. If you are providing code for your paper, and in your analysis you use a complex function from a high-level library, will other researchers be able to edit this function for their own purposes? High-level functions can act as black boxes and any update in the library can result in bitwise reproducibility loss. Reeditable codes were more comprehensible to the majority.

As previously discussed, there are instances where authors may stop using their own software because the sole individual proficient in its operation may no longer be part of the team. This emphasizes the necessity of adhering to robust software engineering principles to prevent such scenarios. Furthermore, in a more typical situation, software may become inaccessible owing to system malfunctions, or the programming language used has become obsolete or archaic that it is impractical for others to adopt it.

## 6 REPRODUCIBILITY PROBLEMS SPECIFIC TO HPC

### 6.1 Parallel computing

Parallel computing involves the simultaneous execution of multiple tasks for efficient processing. Parallel computation introduces new challenges such as out-of-order floating-point arithmetic, which can introduce non-reproducible numerical results (Goldberg, 1991). The combination of nondeterministic behavior in parallel programs and the nonassociativity of addition and multiplication when using floating-point operations poses reproducibility challenges. Minor precision errors owing to the lack of associativity (Figure 3) can quickly influence large-scale computing, which involves billions of operations.

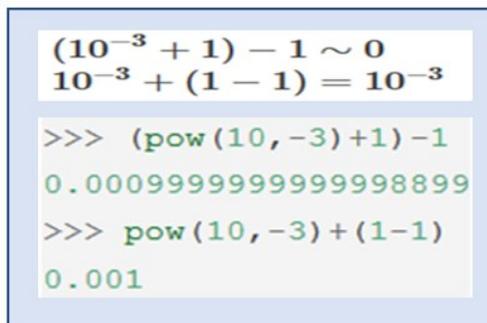

Figure 3: Non-associativity of floating-point operations, an example in Python 3.10.

In (Hunold, 2015), a survey, we learned that a large majority of high performance computing researchers are concerned about the reproducibility of their papers. A large majority (94 %) believed that the reproducibility of articles should be improved in the parallel computing domain. Similarly, they believed that current research articles in the domain of parallel computing are barely reproducible. In parallel computing, the execution order of tasks can vary owing to factors such as task scheduling, load balancing, and multithreading. In (Chohra et al., 2016), the authors suppose that non-determinism can arise from dynamic data scheduling, non-deterministic reductions, resource availability, and different instruction sets.

This nondeterministic behavior can lead to different outcomes, even when the same program is run multiple times under the same conditions. This poses challenges for reproducibility as the results may vary unpredictably, making debugging and result validation difficult. Furthermore, the non-associativity of floating-point arithmetic exacerbates this issue. Floating-point operations such as multiplication and addition are not associative in the fraction space Q, where floating-point real numbers fall. This implies that changing the order of operations can yield different results. When these non-associative operations are combined with parallel computation, achieving bitwise identical results, which can often be a critical aspect of debugging, becomes an even more significant challenge in exascale computing (Demmel and Nguyen, 2013b). Exascale systems, which refers to supercomputers capable of performing $10^{18}$ floating-point operations per second, present immense computational power, but also pose significant challenges for reproducibility. Between the two non-associative floating-point operations, addition is more sensitive and yields different results when computed in different orders. This implies the production of compensation algorithms, such as compensated sums, which are particularly needed for the reduction phase(We present up-to-date solutions in Section 7). With the new frontier supercomputer ([Top500](Top500)) and many other supercomputers that have been proposed for more than a decade, we have exceeded one million parallel computing cores.

The use of grid computing or parallel libraries, such as Message Passing Interface (MPI), with asynchronous message passing on large simulations can frequently lead to out-of-order operation execution. In this case, we do not have repeatability or reproducibility, which drastically increases the difficulty of debugging.

## 6.2 PRNGs and Monte Carlo simulations

In reproducible research on high-performance computing, we have noted many times the statement that Monte Carlo simulations are not deterministic. Although Monte Carlo simulations can be nondeterministic for the same reasons as other computational executions, such as out-of-order floating-point arithmetic and parallelism, we argue that calling Monte Carlo simulations nondeterministic because of the use of a random source can be misleading. To produce scientific results, the Monte Carlo method uses deterministic models of randomness called pseudo-random number generators (PRNGs). This is a scientific approach for precisely mastering the randomness of the reproducibility of each 'independent' experiment. When running stochastic models such as Monte Carlo simulations, experienced scientists use PRNG statuses (simple seeds in the case of old generators). Pseudo-or quasirandom number generation is completely deterministic. The correct use of PRNGs is mandatory for reproducible stochastic computing, particularly when dealing with parallel Monte Carlo simulations. Hellekalek warned simulationists at the 1998 Parallel and Distributed Conference: "Don't trust parallel Monte Carlo" (Hellekalek, 1998). We have also observed frequently the poor advice to initialize your PRNG with "time(NULL)" to enable "true randomness." This is poor advice that should not be applied to science. To enable reproducibility with PRNGs, the initial statuses should be mastered and saved methodically and to select a fine parallelization technique should be selected before running simultaneous "independent" instances of Monte Carlo simulations (Hill et al., 2013). For instance, Figure 4 presents a method for allocating random streams using a random spacing technique, which enables the concurrent execution of autonomous parallel computations. This approach hinges on streams that remain nonoverlapping to maintain their distinct operations. The random spacing method is particularly effective for PRNGs with large periods (e.g., the famous Mersenne Twister family of generators from Mastumoto et al.). The likelihood of overlapping streams is minimal when selecting substreams across the PRNG period. In cases where overlapping might appear (pseudorandom number generators with "small" periods), alternative methods like sequence splitting can be employed, though it is time consuming to pre-compute the initial statuses of non-overlapping streams.

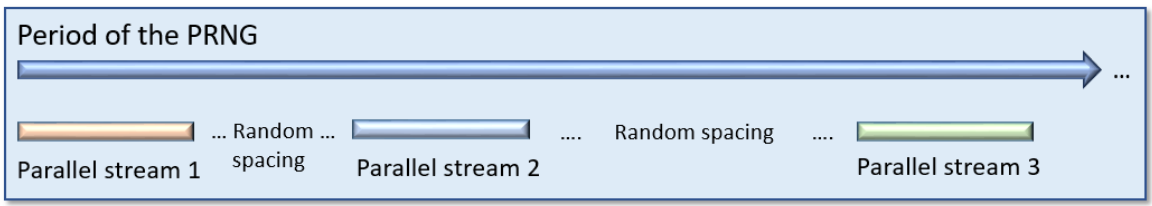

Figure 4: Example of parallel random streams partitioned using random spacing technique.

Certain problems can be encountered while dealing with the parallelization of PRNGs. However, high-quality PRNGs that can be parallelized properly are available now. If the correct methods adapted to the top PRNGs are used, as described in, it is possible to parallelize the Monte Carlo simulations and obtain repeatability with identical bitwise results for all stochastic experiments (Hill, 2015). Here is a short list of high-quality PRNGs used for parallel computing from the selection made by (Antunes and Hill, 2023):

- Philox and Threefry were proposed by Salmon et al. at the 2011 SuperComputing Conference (Salmon et al., 2011). They rely on cryptographic techniques such as Advanced Encryption Standard (AES) (Rijmen and Daemen, 2001). They proposed a sound and easy parameterization technique to solve the problem of distributing "independent" stochastic streams within parallel applications. Because of its cryptographic constraints, it can be considered a slow generator; however, statistically, this type of generator is very good, even if we notice certain reproducibility problems (Hill, 2015). Cryptographically secure PRNGs are known as CS-PRNGs.
- MRG32k3a, a combined recursive generator, was handpicked by L'Ecuyer with brute force to satisfy the most stringent statistical tests developed by (L'ecuyer, 1999). The generator facilitated numerous parallel streams and assets for parallel computing. However, it can be up to 19 times slower than the Mersenne Twister in terms of C/C++ implementation, making it less appealing for intensive parallel computing. MRG32k3a had the highest statistical quality, whereas MT offered a better performance with slightly less statistical soundness.
- WELL, developed by Panneton, L'Ecuyer, and Matsumoto (Panneton et al., 2006), is an improvement upon the Mersenne Twister from a statistical perspective. Originating from linear feedback shift registers (LFSRs), WELL has not been used widely as parallelization technique MT.
- The PCG, a recent PRNG, was created in 2014 by O'Neill (O'neill, 2014). It is claimed to have superior statistical properties compared with other generators. The Numpy documentation recommends this for general purpose because of its qualities.
- The Mersenne Twister, introduced by Matsumoto and Nishimura (Matsumoto and Nishimura, 1998), quickly became known because of its long period of $2^{19337} - 1$. This was the first family of generators designed for GPUs and field-programmable gate arrays (FPGAs). The SFMT version, which uses the vector possibilities of modern processors (single instruction, multiple data), is faster than the original MT. It has better statistical properties, and proposes an even larger period of up to $2^{216091} - 1$; however, it is much less known than the original MT (Saito and Matsumoto, 2006). However, none of the PRNGs in this family is suitable for cryptographic applications.

True random numbers are another family of random numbers. True random numbers are generated in an unpredictable and non-deterministic manner. A true random number generator (TRNG) is a device used to generate true random numbers. Unlike pseudo-random number generators (PRNGs), which use deterministic algorithms and an initial seed to produce sequences of random numbers, TRNGs rely on unpredictable physical processes or sources of randomness to generate truly random numbers. Common sources of randomness in TRNGs include electronic noise, radioactive decay, atmospheric noise or optical noise. To perform reproducible science with TRNGs, each random number must be saved. When dealing with heavy Monte Carlo simulations, such as in high-energy physics, using TRNG is not a suitable option because saving billions of numbers would consume many resources. In addition, true random numbers are often slow to generate compared with pseudo-random numbers. Another category of random number generators is known as quasi-random number generators (QRNGs), which are designed to generate sequences of numbers that approximate a uniform distribution more evenly than true random sequences, whether produced by a TRNG or simulated by a PRNG.

### 6.3 Optimization

Optimization is an important concept in high-performance computing. When we discuss optimizing the execution of a process on a CPU, we might consider the –O2 and –O3 compilation options (for C, C++, and Fortran), and we can also consider fused multiply–add (FMA), advanced vector extensions (AVX), and tensor cores that are able to achieve small matrix multiplications in one cycle. At the lower level, we also have an FMA, which is an arithmetic operation that combines multiplication and addition in a single instruction. The fused nature of this operation means that the multiplication and addition are performed inside the processor at hardware level in a "single" step, improving computational efficiency of the following two operations: a * b + c). AVX is an extension of the x86 instruction set architecture that allows for single instruction, multiple data (SIMD) operations. SIMD enables the processor to process multiple data elements in parallel with a single instruction, thereby increasing the computational throughput. AVX introduces wider vector registers (e.g., 256 or 512 bits), allowing a single-vector instruction to operate simultaneously on multiple data elements. This is particularly beneficial in data-parallel applications, where the same operation must be performed simultaneously on multiple data points. For example, using AVX, a 256-bit vector register can hold eight single-precision (32-bit) floating-point numbers or four double-precision (64-bit) floating-point numbers. AVX2 and AVX-512 further extend this capability with wider vector registers and additional instructions. When compiling with –O3 aggressive optimization, all of these optimization features might lead to a loss of reproducibility. We can even lose run-to-run repeatability on supercomputers as described by Prof. Thomas Ludwig, director of the DKRZ in one of his presentations (). With the generalization of dynamic executions inside CPUs (also known as "out-of-order"), created to better feed the pipeline of CPUs, we observe more issues with floating point arithmetic in HPC. In addition, as shown in (Mytkowicz et al., 2009), even the performance evaluation of an optimization feature, such as-O3, can be non-reproducible.

### 6.4 Hardware heterogeneity

In a high-performance computing environment such as the computing grids used by CERN, hardware resources can vary widely. This heterogeneity can have significant implications for reproducibility in HPC because it introduces variations in performance, precision, and execution behavior across different systems (Boyer, 2022). Indeed, the potential heterogeneity in hardware platforms and networks raises concerns regarding reproducibility. The dynamic scheduling required to adapt to changing resources and loads makes it difficult to consistently execute operations in the same order across different runs on a distributed-memory platform. Figure 5 illustrates a standard workload scenario of the Large Hadron Collider beauty experiment (LHCb) at CERN. In this figure, various worker nodes are engaged in retrieval and processing tasks. These nodes are equipped with a diverse array of hardware components, thereby exemplifying the concept of grid computing in high-performance computing environments.

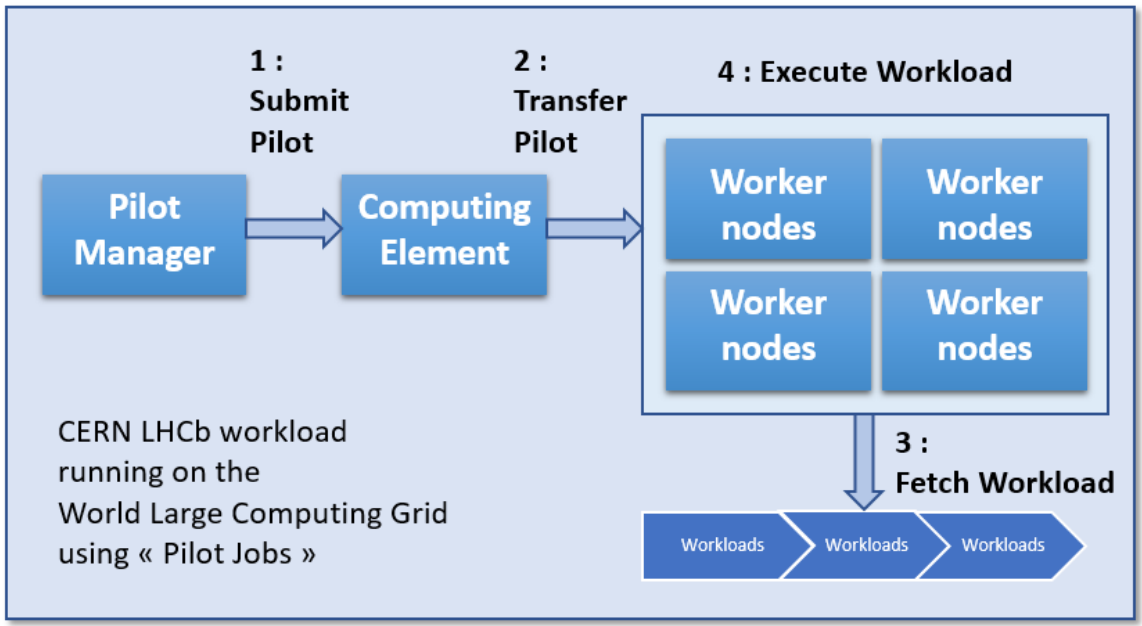

Figure 5: Typical HPC workload for LHCb experiment

We can also encounter performance variability owing to hardware heterogeneity. Different hardware architectures have varying processing capabilities, such as clock speed, number of cores, memory bandwidth, BIOS configurations, and cache size. Consequently, the same code executed on different machines can exhibit different performance characteristics. This variability can lead to discrepancies in the time required to complete the computations, making it challenging to consistently compare and reproduce the results across diverse hardware setups.

In addition, hardware such as CPUs can be widely different, implementing different technologies and instruction sets, including SIMD (e.g., AVX), which can accelerate certain computations. However, code utilizing specific instruction sets may produce different results for CPUs that lack the support for these instructions. This discrepancy can result in performance differences and, in certain cases, numerical divergence of the output. This can also affect floating-point precision (e.g., double precision and extended precision). The use of different precision values can lead to variations in the numerical accuracy and precision of the results. In some cases, these differences may be negligible. However, when repeated intensively, they can significantly affect the final output.

In addition to handling hardware, memory can vary, and with it, their performance. Diverse memory hierarchies, such as different cache sizes and memory access latencies, can affect the efficiency of memory-intensive applications (applications with a memory-bound profile). This can also result in a varying performance.

In computing grids, we are seeking not only parallelization but also load balancing, and this aspect is challenging. Code optimization for a specific parallel architecture may not efficiently utilize other architectures, leading to different parallelization patterns and, consequently, different performances and results.

Finally, software stacks such as compilers, libraries, and operating systems can be different. All these variations can lead to different generated or executed codes, potentially altering the performance and numerical behavior of the application.

Therefore, platform-specific bugs may be encountered. Such issues can lead to nonreproducible results in certain systems, making it challenging to identify and fix the root cause.

**6.5 Quantum computing**

Quantum computing has emerged as a novel branch of high performance computing. This is promising for the optimization of problem solving (Hogg and Portnov, 2000). Tools are being developed to facilitate quantum computing, which may remain obscure for those used in classical deterministic computer science. (Shaydulin et al., 2021) proposed a Python toolkit called QAOAKit designed for research on the quantum approximate optimization algorithm (QAOA). It serves as a unified repository of pre-optimized QAOA parameters and circuit generators for quantum simulation frameworks. By incorporating known parameters for the Max-Cut problem and providing conversion tools for various quantum simulation frameworks, QAOAKit facilitates the reproducibility, comparison, and extension of research results in quantum optimization, thus addressing open questions regarding algorithm performance and behavior.

IBM has been at the forefront of providing access to quantum machines through the quantum experience. This platform allows researchers and generalists to run experiments on IBM's quantum processors, which can have up to 127 qubits. This remarkable access was facilitated by Qiskit, an open-source quantum computing software development framework that enables users to write quantum algorithms using Python. The framework also includes a quantum assembly language (QASM), which provides a method to express quantum circuits at a lower level, allowing for more detailed control over quantum operations.

Furthermore, methods for creating qubits vary and are a topic of active research. In general, qubits can be realized using several physical systems. The most prominent methods include trapped ions, where individual ions are confined and manipulated with electromagnetic fields; superconducting circuits that exploit the quantum mechanical properties of macroscopic electronic circuits cooled to near absolute zero; and semiconductor qubits, which are fabricated using principles similar to those used in conventional computer chips. Each of these methods has advantages and disadvantages, particularly in terms of scalability, coherence time, and error rate. Researchers are continually seeking to improve these systems to create more reliable and long-lived qubits that can pave the way for scalable quantum computing.

However, with the use of quantum computing for high-performance optimization, new reproducibility challenges have arisen. Reproducibility in quantum computing is one of the main concerns, and thousands of excellent qubits are required to obtain a single 'perfect' qubit. In (Mauerer and Scherzinger, 2022), they remain at a high level, proposing solutions to manage packages and dependencies, which is not a problem for quantum computing alone. If we get deeper into the technical aspects of quantum computing, we can indeed understand that it has intrinsic problems (Dasgupta and Humble, 2022) (Dasgupta and Humble, 2021a) (Dasgupta and Humble, 2021b) (Hill et al., 2023) owing to the non-deterministic behavior and the fact that it is physically difficult to have reliable qubits due to decoherence, noise, and lack of error correction. IBM offers free access to small quantum machines; thus, considerable research has been conducted on these machines. In (Hill et al., 2023), they implemented a three qubits Grover algorithm. The oracle is coded (with quantum gates) to find the property "being equal to five." They executed the algorithm on a quantum simulator, showing that they should obtain the right solution with a probability of 94%; quantum computing is essentially stochastic, and the results are given by probabilities. However, when run on different real quantum machines, the results were not good, and we found considerable variability. At most, we have a correct solution of 60% (which is much less than what was expected according to the simulation, but is still usable). The reproducibility problem occurred when running this experiment on the same quantum machine several times, or on different quantum machines, with exactly the same code and parameters. Statistically different results were obtained each time, and most of them were not the expected results. This questions the usability of current quantum computers(B. A. Antunes et al., 2024). Simulators or hybrid quantum machines are used to develop skills; however, we cannot produce thorough scientific results without reliable reproducibility.

Quantum computing remains a young technology, and it is probable that we will encounter disruptive technologies and obtain general quantum machines with perfect qubits over the next few decades.

## 6.6 Machine learning

One recent use of high-performance computing resources is in machine learning and big data. Machine learning uses randomness to train models. However, practices with good reproducibility are not widespread. It can be difficult for scientists to obtain repeatable pseudorandom number sequences using high-level frameworks (https://neptune.ai/blog/how-to-solve-reproducibility-in-ml). Furthermore, deep-learning frameworks undergo updates and version changes, which can introduce discrepancies when comparing the results obtained with different software versions.

Machine learning is often associated with big data that can present challenges to reproducibility, primarily owing to dataset or data-sampling variability. As the size of the datasets increases, filtering and sampling techniques become increasingly important. Different sampling approaches can lead to variations in the data subsets used for analysis, which can potentially affect the reproducibility of the results. When dealing with classification, the exact numerical results are much less important. However, parallel processing and distributed systems, which are commonly employed in big data analytics, can contribute to nondeterministic behaviors, making reproducibility challenging. Parallel algorithms may exhibit different behaviors based on factors such as the execution order or allocation of computational resources. Consequently, obtaining the same results across different parallel executions for debugging purposes can be challenging. In addition, when dealing with massive quantities of data, it becomes more difficult to share such data, which affects the ability of other researchers to reproduce the results. With (Collberg and Proebsting, 2016), we showed that research studies financed by public institutions are more easily reproduced because code sharing is a more common practice. Private companies have a heavy influence in the world of machine learning and are less inclined to share their code and data because they compete to obtain industrial advantages.

Finally, several computing resources using graphical processing units (GPUs) have been proposed. This hardware is widely used in the AI field, particularly for training algorithms, because modern GPUs embed a large number of tensor cores. Reproducibility can lead to several challenges when GPUs are used in high-performance computing. Floating-point precision is less important, and training can be achieved with half precision (FP16) or sometimes even less. However, nondeterministic rounding errors may occur during numerical computations. This can lead to slight variations in the results even when the same code and data are used. Furthermore, different GPU architectures may yield inconsistent results, making reproducibility across diverse hardware configurations challenging. GPU libraries and drivers also contribute to the reproducibility challenges. Several optimization libraries in machine learning and big data are vendor-specific and rely on proprietary interfaces and functionality provided by GPU manufacturers. Incompatibilities or differences between library versions or GPU drivers can introduce variations in the results and hinder reproducibility. Parallel execution on GPUs can also introduce "nondeterministic" behaviors that further affect reproducibility. Interference between GPU kernels, variations in thread scheduling, and race conditions during parallel execution can lead to different outputs across runs, even with the same input data and code (Jézéquel et al., 2015) (Taufer et al., 2010).

## 6.7 Silent errors and other soft errors

Not often mentioned, silent errors has a significant role in the reliability of HPC. Disturbances of either electrical or magnetic nature within a computer system can lead to an individual bit in dynamic random access memory (DRAM) undergoing an unplanned switch to the opposite state. Initially, this phenomenon was primarily attributed to alpha particles emitted by impurities within the chip packaging. Alpha particles or alpha radiation consist of two protons and two neutrons, which are generally emitted during the process named alpha decay, but may also be produced in other ways (Rutherford and Royds, 1908). However, (Normand, 1996) demonstrated that ambient radiation is the primary cause of isolated and temporary errors in DRAM chips. Neutrons stemming primarily from cosmic ray secondary interactions have been identified as a key source of this radiation, capable of altering the contents of one or more memory cells or disrupting the circuitry responsible for their read or write functions. An extensive study on DRAM was attempted to determine how common memory errors are (Schroeder et al., 2009). In this study, the behavior of DRAM errors in real-world scenarios deviated significantly from popular assumptions. Notably, the observed DRAM error rates are dramatically higher than previously assumed, ranging from 25,000 to 70,000 errors per billion device hours per Mbit, and over 8% of the dual inline memory modules (DIMMs) experience errors annually. Contrary to the earlier beliefs, it has been demonstrated that hard rather than soft errors dominate the landscape of memory errors. Temperature, which is a known influential factor in controlled environments, exhibits a minimal impact on error behavior in practical settings when accounting for other variables. CPUs cannot avoid this problem. In advanced computing, growing surface areas, shrinking fabrication sizes, and higher component densities have led to an increase in the occurrence of bit flips, which can cause silent errors. Although such anomalies are believed to occur more frequently in DRAM, other components, such as logic gates and arithmetic units, can also be vulnerable (Elliott et al., 2013). This study aimed to assess the impact of a single-bit flip on specific floating-point operations. The analysis delves into errors resulting from flipping particular bits in the IEEE floating-point representation, avoiding reliance on proprietary information, such as bit flip rates and vendor-specific circuit designs. In (Dixit et al., 2021), they are also considering CPUs to generate hard errors. They argue: *"For example, when you perform 2x3, the CPU may give a result of 5 instead of 6 silently under certain microarchitectural conditions, without an indication of the miscomputation in system event or error logs. As a result, a service utilizing the CPU is potentially unaware of the computational accuracy and keeps consuming the incorrect values in the application."*

From these results, we know that DRAM and CPUs can lead to silent errors. As technologies evolve, working with smaller chips, increases the probability of a bit flip. At the exascale, we have supercomputers such as Frontier that have 8 699 904 cores. At this scale, the mean time before failure (MTBF) decreases to a few hours for safe computing. In the following section, we discuss the proposed solutions.

## 7 WHAT ARE THE SOLUTIONS PROPOSED TO ENHANCE REPRODUCIBILITY?

### 7.1 Versioning and archiving

Ensuring the reproducibility of the computational aspects of a paper requires versioning, archiving, documenting, and sharing code. Versioning is important for codes and scripts. Consider a scenario involving a computer experiment that produces interesting figures. Then, you will continue to update your code. Later, you want to work again on these figures, perhaps because you submitted a research paper and the reviewers asked you, or maybe it is just in your research process that you want a deeper understanding of the figures produced. Because you have modified the code, you are now unable to generate the same figure, you are not getting the same interesting results, and you do not have the original version of the code. If this sounds strange to serious software developers with code versioning, this situation is not uncommon, and helps them understand the need for tools to control code versions. Second, it is crucial to permanently archive the codes. If the Html error code "404 not found" appears more times than expected, we have to pay attention to the place where we store our code versions. Third, you need to document your code so that other people can use and understand it. Finally, you need to share your code so that people can access it (Kitzes et al., 2018).

The two prominent versioning tools are Git and Apache Subversion. Git has emerged as the dominant tool in this regard and is widely used in versioning. Although Git offers numerous features for reproducible research, we advocate its integration into the development process, acknowledging that beginners may require an initial learning phase. Essentially, adopting Git represents the basis for best practices.

Several platforms are available for archiving codes. GitHub and GitLab were the most prevalent. They also function as repositories. When developing code for a research project, the use of one of these tools, along with Git, is mandatory. However, these tools have not been initially used in reproducible studies. It is worth noting that these platforms, as commercial entities (in the case of GitHub), do not guarantee perpetual code accessibility. Alternatives, such as Zenodo and Software Heritage, offer more robust solutions for code archiving (Hinsen, 2020). Zenodo and Software Heritage developed by CERN and Inria respectively, provide a means of archiving code, data, or any numerical resource associated with an article, each identified by a unique ID (either DOI or SWHID). Zenodo is more oriented toward archiving the code and data of an article, whereas Software Heritage is a larger project that stores all existing versions of open-source code in the world, such as libraries and paper artifacts.

**7.2 Literate programming and documentation**

To solve the documentation problem (make the computation as clear as possible), and also maybe a bit to solve the workflow problem, literate programming was introduced. When sharing codes, it is essential to make them accessible (for ease of use). The concept of literacy programming emerged in 1984 with (Knuth, 1984). The principle of this concept is to intertwine codes with natural language text to make the entire content more understandable. The current application of this principle is in the form of notebooks (computational documents), which are interactive documents that allow text to be combined with code blocks. They are available in various forms and languages. To facilitate the writing of this type of file, languages for text composition are used, primarily LaTeX and Markdown. The latter is much lighter and more comprehensible without an interpreter, as discussed by (Pouzat, 2021). Knuth developed the initial tool for literate programming called WEB, which comprises two main programs: Tangle and Weave (Knuth, 1984). This system was tailored to the Pascal programming language and generated documents formatted in TeX. Later, Knuth and Levy (Knuth and Levy, 1994) created a version of the C language called cweb. The contemporary evolution of these tools is noweb (Johnson and Johnson, 1997), which is designed to be adaptable across languages. Its central programs, notangle and noweave, are both coded in C. Documents produced through noweave can be formatted using TeX, LaTeX, or troff, or even displayed within a web browser as HTML. Software utilities, such as WEB, cweb, and noweb, empower authors to craft both written content and code; however, they lack mechanisms for executing code directly within documents. Instead, the code destined for execution is extracted, resulting in source code files that are forwarded to a compiler or interpreter. Perhaps the most widely adopted tool for achieving reproducible research is Sweave, which is extensively used in the R programming community. The Sweave methodology for reproducible research has led to analogous tools such as SASweave and Statweave. Some of these tools cater to statistical languages other than R and are compatible with document preparation systems other than LaTeX, encompassing formats like Open Document Format and Microsoft Word (Lenth and Højsgaard, 2007) (Baier and Neuwirth, 2007) (Lenth, 2012). However, Sweave and its derivatives lack the capability to rearrange code blocks during tangling. Consequently, their support for literate programming remains limited. Among the most well-known literate programming tools, we mention the following:

- Jupyter (Kluyver et al., 2016), a notebook primarily used in Python.

- RStudio (Allaire, 2012), developed mainly for the R language.
- Org-mode (Schulte et al., 2012), a more versatile tool that can be used with all languages.

These three tools are extensively presented in detail in the Inria Reproducible Research MOOC (https://www.fun-mooc.fr/fr/cours/recherche-reproductible-principes-methodologiques-pour-une-science-transparente/). The use of notebooks and, more broadly, literate programming is highly endorsed by the scientific community. (Stanisic et al., 2015) proposed a workflow solution for reproducible research based on the use of the Git and org-mode. (Desquilbet et al., 2019) discussed the use of notebooks in their work. (Stodden et al., 2013) compiled a list of tools for reproducible research, with notebooks as an evident component. (Ragan-Kelley et al., 2018) offered a tool named Binder, which allows Jupyter notebook repositories to be directly executable on the web without installation by linking to a Git repository. (Delescluse et al., 2012) also mentioned the org-mode and Sweave for reproducible research within the context of R language.

With the increasing use of the Python programming language, Jupyter is probably the most commonly used and easy-to-use notebook. However, owing to its versatility, org-mode is probably the most plebiscite tool, particularly for enthusiasts of Emacs, because it uses this text editor (Stallman, 1981).

Large, complex software applications cannot be fully coded using literate programming. In this case, literate programming (with current tools) was used to run or analyze the results. Therefore, to deal with such an application, other documentation tools are required, rather than just literate programming. There is not only one way of creating good documentation (Sommerville, 2001), but good software engineering practices are needed. One well-known method to document object-oriented code is to produce corresponding diagrams using the unified modeling language (UML), for instance (Booch et al., 1996).

In addition to literate programming and comprehensive documentation, adhering to the FAIR principles (findable, accessible, interoperable, and reusable) can help in enhancing reproducibility in scientific research. These principles advocate the creation of digital assets, such as data, algorithms, and tools. Implementing these principles ensures that research outputs are discoverable through unique identifiers and rich metadata accessible via well-defined protocols, interoperable through standard formats and vocabularies, and reusable through clear usage licenses and provenances. The use of FAIR principles facilitates the replication of research results and fosters collaboration and innovation by enabling researchers to easily share and build upon each other's work (Wilkinson et al., 2016).

### 7.3 Workflow

Scientific workflows are an important part of reproducible research. It is essential to understand the sequence in which operations should be executed. One of the first tools to emerge was Makefile. This allowed for automation of the executable construction. The first paper to consider this for this purpose appeared in 2000 (Schwab et al., 2000), with Jon Claerbout, who was one of the early contributors to reproducible research in 1992. Schwab et al. described how to use "make" and related tools along with naming conventions for input and generated files to ensure that readers as well as authors can reproduce computations. With (Cohen-Boulakia et al., 2017), biology is one of the fields that carry the most workflow tools to enhance reproducibility, perhaps because biology produces and analyzes a large amount of data. In (Oinn et al., 2004), it is given: "*in silico experiments in bioinformatics involve the coordinated use of computational tools and information repositories. A growing number of these resources are being made available with programmatic access in the form of Web services. Bioinformatics scientists will need to orchestrate these Web services in workflows as part of their analyses*". A lot of workflow tools have been developed since the creation of the Makefile in 2000 (more than 300) (Amstutz et al., 2024). These aim to be easy to use and often provide graphical user interfaces (GUI). Taverna (Oinn et al., 2004) was an important open-source scientific workflow management system that provides a workbench for designing workflows and a workflow engine for execution. Taverna optimizes the workflow structure, links web services and activities, and supports the execution of grid and cloud computing resources. However, the project is now retired. Galaxy (Giardine et al., 2005) is a popular workflow system in the bioinformatics community that recently have switched from an XML based serialization to a YAML/JSON serialization. While Galaxy has been working with Common Workflow Language (CWL), a format for workflow specification (Amstutz et al., 2016), CWL is not currently supported. The CWL project is defined in (Crusoe et al., 2022). Galaxy provides a GUI for exploring and sharing the workflow execution. Galaxy manages the activity dependencies for parallelization, generates and monitors tasks, and employs dynamic scheduling for task dispatch. OpenAlea (Pradal et al., 2015) is an open-source scientific workflow system linked to a Python modeling tool. It allows users to export workflows into CWL format and provides provenance tracking. However, it has limitations in terms of complexity, lack of a central repository, and difficulty visualizing provenance data. NextFlow (Di Tommaso et al., 2014) is a command-line-based workflow system for complex parallel scientific workflows. It uses text-based specifications, and supports processors and operators for workflow execution. It has limitations in terms of managing tool dependencies, querying provenance data, and a lack of structured provenance information. Pegasus (Deelman et al., 2004) is widely used in multiple disciplines, and has features such as portability, optimized scheduling algorithms, scalability, provenance support, and fault tolerance. It consists of five components–a mapper, local execution engine, job scheduler, remote execution engine, and monitoring component–to generate an executable workflow and schedule its execution on different infrastructures. Similar to Pegasus, Swift (Zhao et al., 2007) is used in various disciplines and focused on data-intensive workflows. It executes workflows through phases, such as program specification, scheduling, execution, provenance management, and provisioning. This supports workflow partitioning, fault tolerance, and dynamic resource provisioning at various execution sites. Kepler (Altintas et al., 2004) is a workflow management tool that allows for different execution models in workflows. It has a graphical workbench that supports actors in the workflow steps. Kepler offers static or dynamic scheduling, supports fault-tolerance mechanisms, and executes workflows on web services, grid-based actors, or Hadoop (Borthakur, 2007) framework. Chiron (Ogasawara et al., 2013) used a database approach to manage the parallel execution of data-intensive scientific workflows. It uses algebraic data models and operators to express data and workflow activities. Chiron supports workflow monitoring, parallelism (data, independence, and pipelines), and dynamic scheduling. It stores execution and provenance data in a structured database. Triana (Taylor et al., 2007) is a GUI tool that was initially developed for data analysis in the GEO600 project. Triana was designed to manage distributed applications. Finally, Snakemake is a powerful workflow engine that provides a readable Python-based workflow definition language. It provides an execution environment that can scale from a single-core execution to full parallel computing on clusters with several cores without modifying the workflow. This tool is widely used in bioinformatics (Köster and Rahmann, 2012).

Taverna, Chiron, Triana, and Galaxy have features such as a GUI for workflow design, support for independent parallelism, dynamic scheduling, and workflow execution in grid and cloud environments. Taverna supports the sharing of workflow information, whereas Galaxy specializes in executing bioinformatics workflows. Sequanix proposed a dynamic graphical interface for Snakemake (Desvillechabrol et al., 2018). An almost complete state-of-the-art on workflow system management is available in (Liu et al., 2015), and a review more focused on life science was published in 2017 (Cohen-Boulakia et al., 2017). (Ivie and Thain, 2018) have proposed a more general study on workflow management.

## 7.4 Software environment

*7.4.1 Handling dependencies*

When writing software, coding for data analysis, etc., you will always have to rely on your software environment. Finally, when publishing the results, other researchers may face difficulties. It is possible that they cannot reproduce their results because they do not have the same software environment used originally. Having good documentation is neither sufficient, nor does it describe the workflow or version of history.

Several solutions exist to address these dependencies. For example, the use of "apt-get" on Linux distributions, or the use of "pip" when working with Python. The command-line tool, apt-get is a package management system that enables the easy installation, upgrade, and removal of software packages. You can install specific versions of packages using Snapshot. It automatically resolves dependencies, ensuring that all required software libraries and components are installed. The command-line tool pip is a package installer for Python that simplifies the process of managing Python libraries and their dependencies. This allows users to easily install, upgrade, and remove Python packages from the Python Package Index (PyPI). However, both require manual installation of packages (even if they can work by themselves on the dependencies between packages). Even if such tools are suitable for use alone, they are not suitable for reproducible research. Python has been adopted for numerous scientific prototyping computations. There are certain specific tools for Python such as Conda, which are packages and environment managers commonly used in the Python scientific computing community. Conda allows for the installation of both Python and non-Python packages, making it useful for managing complex software environments with dependencies beyond Python libraries. Finally, more advanced features allow the creation of a virtual environment. Venv is a tool that helps to create isolated Python environments. This allows users to create separate environments for different projects, thereby preventing conflicts between dependencies. When a new virtual environment is created, it comes with its own copy of the Python interpreter and a local copy of the pip installer. This ensures that any library installed in the virtual environment is isolated from the global Python environment, thus making it easier to manage and reproduce project-specific dependencies. Venv is particularly useful when working on multiple projects or testing different versions of libraries.

While tools such as apt-get, pip, Conda, and venv are widely used and provide convenient ways to manage software environments and dependencies, they have limitations compared to more specific tools designed for reproducibility. First, as they are not designed for reproducibility, the stability of software environments can be challenging to achieve with these tools because the exact versions and configurations of packages may not be explicitly captured. This can lead to loss of reproducibility. For example, if you are using a library such as Pandas or Numpy in Python, developers might change certain inner code without changing the version number; therefore, tools such as pip (and those based on it) will not clearly identify the two versions of the code. Another weakness of such tools is that they often require administrative privileges to install packages system-wide, which is not suitable in an HPC environment where there are no root privileges. venv provides a level of isolation but is limited to Python libraries.

Therefore, some tools have emerged to provide satisfactory solutions for reproducing the computational aspects of a research paper. CDE (Guo and Engler, 2011) is a tool designed to address dependency issues. It packages code, data, and the environment. Root access is not required. The tool works as follows: it utilizes ptrace and automated packaging of code, data, and environment to execute on a Linux machine. A known issue is that it may not include all dependencies that must be added manually. In addition, it operates only with compatible Linux kernels. Here is how it operates: Alice initiates the command "cde script.py data.dat". This action will execute the Python script, capturing information through ptrace, and creating a "cde-package" directory. Alice compresses this directory and sends it to Bob. Bob then executes the command "cde-exec script.py data.dat". Instead of searching for libraries in Bob's PATH, cde-exec employs ptrace or strace to modify library usage and utilize the ones stored in the folder that Alice sent. There is a potential performance loss ranging from 2 to 28%. This solution may not be suitable for HPC because of the potential loss in performance. Sumatra tools (Davison, 2012) enhance reproducible research by providing systems for automated metadata/context capture. The core library of Sumatra implements functionalities such as capturing the hardware and software environment, input and output data, and scientific context. This core functionality can be used by different interfaces that cover various working methods (command line, graphical interface, etc.). Sumatra tools do not necessarily need to capture every piece of information to be effective; the more information recorded, the greater the ease of reproducing results. These tools are designed to be easy to use and do not slow down the scientist's usual workflow. Sumatra tools consist of a core library implemented in Python, a command line tool, and a web browser-based interface. The command line tool allows capturing the computational context, inputs, and outputs, as well as viewing previous computations. The web browser-based interface provides additional capabilities for viewing, searching, and annotating computation records. Similar to CDE, ReproZip (Chirigati et al., 2016) is a tool that aims to make computational experiments reproducible across different platforms long after they are created. It automatically captures the provenance of an experiment by tracing system calls using ptrace, and uses this information to create a lightweight reproducible package that includes only the files required for reproduction. ReproZip adds a feature that is compatible with Linux, Windows, and macOS. Its limitations are related to distributed environments, such as MPI or Hadoop clusters. However, neither Sumatra nor ReproZip provide a performance analysis similar to that of CDE. Because ReproZip uses ptrace as CDE, a plausible hypothesis is that the performance should be similar. Ultimately, these tools are used for reproducible research, but are probably not suitable for high-performance computing.

As the handling of such tools requires a learning phase, the final objective is to find a tool that can be used by everyone and is suitable for high-performance computing. Guix is the recommended tool for handling library dependencies (Courtès and Wurmus, 2015). Guix is a software package manager that uses a functional paradigm to manage software dependencies. In this paradigm, the build and installation processes are considered as pure functions, meaning that their results depend solely on the inputs they receive. This approach allows for the effective caching of the results on the disk, ensuring that identical inputs will always yield the same results. Implementation of this functional paradigm involves strict control of the build environment. Guix, along with the Nix package manager, from which it draws inspiration (Dolstra et al., 2004), uses a privileged daemon to build processes in Linux containers, specifically in the chroot environment. These containers have their own user ID and separate name spaces for process IDs, inter-process communication, networking, etc. The chroot environment includes only explicitly declared input directories, preventing the build process from accessing unintended tools or libraries. Separate name spaces also ensure that the building process does not communicate with the outside world. The results of each build process are stored in a common directory called the "store," usually located at /gnu/store. Each entry in the store has a name that incorporates a hash of all the inputs used in the corresponding build process. This includes not only compilers and libraries, but also build scripts and environment variables. The hash calculation considers the complete dependency graph for each package, recursively including the directories of the tools and libraries used during the build. This deterministic approach enables reproducibility and the system can access the entire dependency graph used to build each package.

We found that Guix is probably the best tool for handling dependencies. In addition, this tool is not only providing a "binary" unreadable file to execute to reproduce any computation (but in this case, we do not really know what we are executing), but also it provides a clear and complete vision of the software environment being used. Each version was identified by a unique hash, with its source available in the Guix repository. A clear and easy to use tutorial proposed in 2023, enables the production of reproducible research papers (available at https://hpc.guix.info/blog/2023/06/a-guide-to-reproducible-research-papers/), and we recommend its reading. The downsides of Guix are that it is not widespread in computing clusters, and that daemon running needs root privileges; hence, the first setup and installation of Guix need to be done by a system administrator of the cluster. When you are working locally with Guix, it allows you, if future users do not have Guix, to generate a Docker or an Apptainer image (evolution of Singularity). Hereinafter, we present these two tools.

*7.4.2 Virtual Machines*

To make computational research reproducible, another existing tool is the virtual machine (VM). One of the first detailed mentions of this concept arrived 50 years ago (R. P. Goldberg, 1974). Using modern virtual machines, we employ a hypervisor as a virtualization platform that allows multiple operating systems to run concurrently on a single physical machine. Currently, there are two types of hypervisors, as described in Figure 6, native and hosted. Native hypervisor, also known as "bare metal," is a software that runs directly on hardware; the native hypervisor interacts directly with the host's hardware, providing a platform upon which multiple operating systems can be run by the hypervisor. This type of hypervisor is a lightweight and optimized host kernel. On the contrary, hosted hypervisor is a software that runs inside another operating system. Therefore, the guest operating system will run at a third level above the hardware.

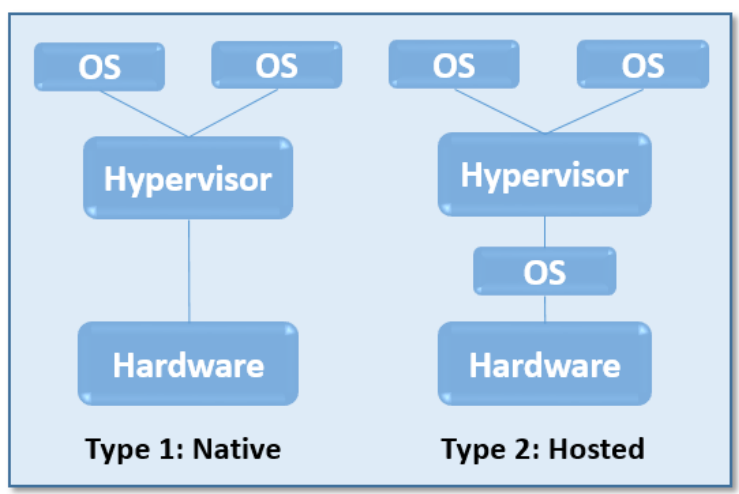

Figure 6: Hypervisor Type 1 and Hypervisor Type 2.

Within a virtual machine, there is an operating system dedicated to your experiments and you are able to "capture" everything. This enabled other researchers to rerun the experiment inside the virtual machine. The disadvantage of this solution is that it is costly in terms of performance, which prevents scientists from running computation-intensive applications to run them. However, this is of interest for most applications. In (Stodden et al., 2013), they mentioned VirtualBox and VMWare as the VM tools used for reproducible research. These were the same mentioned in (Ruiz et al., 2014). These two tools use the second type of hypervisor: one that is much easier to use (ease of use is a strong criterion for reproducible research) and one that may cause more overhead (which is not suitable for high-performance computing). Some scholars have studied the performance of Type 1 virtualization, and few others have studied Type 2 virtualization, often in the context of cloud computing, where this kind of technology is ubiquitous in providing usable operating systems for many people based on the same hardware (the main purpose is not reproducible research). In (Gilbert et al., 2005), they evaluated the virtualization performance in the context of high-energy physics applications at CERN in a grid environment. Their results indicated that virtualization can result in a runtime overhead of up to 15%. They worked on the Type 1 hypervisor tool ESX VMware. In (Matthews et al., 2007), they did not find that virtualization tools such as Xen or VMware generated an overhead, whereas in (Padala et al., 2007), they found out that Xen generated a lot of overhead, possibly due to a higher number of L2 cache misses. For (Acharya et al., 2018), the results are that, on x86 CPUs, a hypervisor (using KVM) performed worse than other technologies. However, this was not the case for ARM CPUs.

All previous studies have evaluated the performance of native hypervisors. We can observe that this is probably not a solution suitable for high-performance computing because of the overhead, but it is also unsuitable for reproducible research because of its complexity. We did not find any reproducible research paper exposing Type 1 hypervisor technologies or tools to make the research reproducible. However, for a hosted hypervisor with VirtualBox and VMware (hosted version), a reproducible research axis was considered.

In a recent paper from 2021, (Đorđević et al., 2021), studied the performance of VirtualBox compared to a native execution, and concluded that "*when it comes to VirtualBox, the performance drop is noticeable*." In (Beserra et al., 2015), they considered the context of high-performance computing, unlike previous studies. They aimed to analyze the performance of KVM (hypervisor Type 1) and VirtualBox (hypervisor Type 2). Three tests (HPL, DGEMM, and FTT) were performed for processing. First, KVM achieved near-native performance and performed 14% better than VirtualBox. Second, KVM has a 5% lower performance than native execution, but performs 10% better than VirtualBox. In the third test (FFT), KVM and native performed 30% better than VirtualBox. Regarding memory access, VirtualBox performed approximately 25% worse than KVM and native. The results were similar for communication. Overall, we can clearly observe that Type 2 hypervisors are not suitable for high-performance computing. A paper merging virtualization, cloud computing, and reproducible research was proposed by Bill Howe (Howe, 2012). He proposed the use of Amazon company services for collaboration in the cloud. However, using private companies as third parties is not always suitable for global open science.

Additional drawbacks of using virtual machines in reproducible research are their opacity and resource intensity. Distributing large VM files across networks can be cumbersome and the VM content remains obscured until execution. Although VMs allow for accurate reproduction, they are not optimal in terms of scalability and adaptability.

*7.4.3 Containers*

Containers are another type of virtualization, often referred to as operating-system-level virtualization. Container and hardware virtualization technologies are similar in that they allow multiple isolated applications to run on a single host computer. However, there are significant differences in how they achieve this goal. Hardware virtualization involves running a complete guest operating system within an isolated environment. This requires a significant amount of resources including memory and storage space. Starting up a guest operating system can take minutes. In contrast, containers share the underlying kernel of the host operating system and isolate the processes running within them from other processes on the host. This implies that containers utilize host dependencies and features, resulting in more efficient resource usage. Multiple containers can be run on a single host and starting a new container is fast, similar to starting a native application. Thus, containers are considered lightweight, standalone executable packages that include an application, all its dependencies, and the share of the host OS kernel. When compared to VMs, this makes them much smaller and more resource-efficient, enabling a quick start with a smaller use of resources. They were initially designed for consistent deployment across various environments from development to production. Containers provide a more efficient and lightweight solution than traditional virtualization methods for running isolated applications.

Containers are currently the leading solution for ensuring reproducibility in computational research. Containers have been cited in several reproducible research papers as the main tool for ensuring the reproducibility of computational experiments. The container encapsulates the application and its dependencies in a self-contained, lightweight, and portable unit. These containers provide an isolated environment to run applications across different operating systems and computing environments, ensuring consistent behavior and reproducibility of results, similar to what was achieved by virtual machines, but in a more efficient manner (Boettiger, 2015).

The first technology studied was Docker (Merkel, 2014). Several studies have advocated the use of Docker for reproducible research (Nüst et al., 2020) (Boettiger, 2015). Since Docker was the first studied technology, it has been compared with hypervisor-based virtualization for sharing software environments. Therefore, many studies have compared Docker virtualization performance and suitability with hypervisor-based virtualization. (Felter et al., 2015) worked with a kernel virtual machine (KVM) (Kivity et al., 2007), which is another Type 1 hypervisor, that is often studied. They applied LINPACK to native Linux, Docker, and KVM. The LINPACK execution consumes most of its time for performing mathematical floating-point operations. The performances were almost identical for Linux and Docker; however, the KVM performance was markedly worse. They performed several different tests, not only pure floating-point arithmetic, and concluded that Docker equaled or exceeded the KVM performance in every case. Chae et al. (Chae et al., 2019) compared Docker with KVM, and the results showed that Docker was faster than KVM, finding that Docker performed better than KVM in terms of CPU, HDD, and RAM. (Rad et al., 2017) also found that Docker performed better than virtual machines. (Chung et al., 2016) found that Docker generates less overhead than virtual machines. (Potdar et al., 2020) compared Docker and virtual machines in terms of CPU performance, memory throughput, disk I/O, load test, and operation speed measurement, and observed that Docker performed better than virtual machines in every test. (Ruiz et al., 2015) also confirmed that virtualization technology using containers performed better.

From this set of studies, we can note that containers such as Docker are better for use than virtual machines in the context of reproducible computational research. However, we did not determine whether Docker induces a performance overhead compared to native execution and if it is suitable for high-performance computing. Although Docker seems to be a good option, it is not suitable for high-performance computing. First, Docker requires administrative privileges. Within high performance computing systems, no user has administrative privileges. Therefore, Docker is unusable in a high performance computing contexts. Second, security concerns have been mentioned in various studies (Priedhorsky and Randles, 2017) (Jacobsen and Canon, 2015) (Zhou et al., 2022).

Other containers have been developed for high-performance computing, such as Charliecloud (Priedhorsky and Randles, 2017), Shifter (Gerhardt et al., 2017), Singularity (Kurtzer et al., 2017) (now renamed Apptainer), Sarus (Benedicic et al., 2019), and Podman (Gantikow et al., 2020). Charliecloud runs containers without requiring root permissions or daemons. It converts Docker images into tar files and unpacks them on HPC nodes. It is considered secure and supports MPI. It solves library compatibility issues by injecting host files into the images. Shifter is an HPC container engine developed by NERSC, which utilizes Docker for image building. It flattens images to ext4 and compresses them into squashfs for storage in parallel file systems. It supports MPI and handles GPU driver compatibility by swapping the GPU driver at container startup. Apptainer (Singularity) is widely used for HPC in academia and industry. It operates with user privileges and does not require a daemon process. It supports GPU, MPI, and, most importantly, InfiniBand, which is a reference in terms of interconnections for HPC systems. Apptainer allows portable execution via a single image file in SIF format. This approach offers hybrid binding models for MPI applications. Sarus is another HPC container engine that relies on a runc to instantiate containers according to the OCI specifications. It uses a bundle comprising a root file-system directory and a JSON configuration file. Sarus can be extended using custom OCI hooks such as MPI hooks. Podman executes containers without privilege escalation by using a user namespace. This supports the same runtime as Sarus and Docker (runc), as well as a faster runtime crun. This study introduces the concept of pod-to-group containers for complex applications. However, they are currently limited by kernel features that are incompatible with network file systems. All of these technologies are based on a basic Linux functionality called Linux Containers (LXC)(Senthil Kumaran, 2017).

Research has shown that none of the operating system-level virtualizations, for example, containers, induce a significant overhead (affect performance negatively) or have at least a near-native performance. (Younge et al., 2017) demonstrated that Singularity operates at the native level for high-performance computing. (Xavier et al., 2013) showed that LXC virtualization offers native performance. A more complete study from 2019 (Torrez et al., 2019) examined the performance of Shifter, Charliecloud, and Singularity. They found no meaningful differences in performance among the four environments, with the exception of modest variations in memory usage. (Yong et al., 2018) studied the performance of Singularity and Docker, and the results showed that Singularity induced no performance loss, whereas Docker induced a slight degradation. (Hale et al., 2017) and (Le and Paz, 2017) confirmed that Singularity does not induce any performance loss. (Abraham et al., 2020) studied the input/output throughputs of Docker, Charliecloud, Singularity, and Podman on a Lustre file system. Results "*shows startup time overhead for Docker and Podman, as well as network overhead at startup time for Singularity and Charliecloud. Our I/O evaluations show that with increasing parallelism, Charliecloud incurs large overhead on Lustre's MDS[1] and OSS[2]. Moreover, we find that the observed throughput of containers on Lustre is at par with containers running from local storage*". (Casalicchio and Perciballi, 2017) found a small overhead for Docker, approximately 5–10% for CPU load depending on the charge, and 10–30% for disk I/O overhead. Focusing on high-performance computing scenarios, (Hu et al., 2019) found no overhead for Singularity in MPI and GPU parallel applications.

Based on this knowledge, we can note that Apptainer (Singularity) and other container technologies, except for Docker, are suitable for high-performance computing and reproducible research. We believe that the Apptainer is the best solution to use and learn because it is the most widespread among computing clusters. However, as Apptainer has recently changed, an up-to-date study could be useful to confirm that Apptainer still remains perfectly suitable with native or near-native performance.

One disadvantage of containers is that they are initially designed for deployment and not for archiving. Although it is possible to create a Docker file that specifies the exact steps for building a container image, there is still no guarantee that the resulting image will be completely reproducible. This means that even if two people use the same Docker file, they may end up with slightly different container images. One reason for this lack of reproducibility is that container images are binary, meaning that they are compiled and are not easily inspectable or modifiable. This makes it challenging to understand all dependencies and configurations within a container image. If someone wants to reproduce the computations performed within a container, the exact container environment must be recreated. However, because of the non-reproducibility of container images, this process may not yield identical results. Therefore, although containers are efficient for deploying applications, they may not be ideal for achieving full reproducibility in scientific computing or research contexts. Containers are the most commonly used tools in practice and are suitable for reproducible research. As mentioned previously, Guix is currently the best solution for achieving computational reproducibility. Complete state-of-the-art containers for high-performance computing were published in 2022 and 2023 (Zhou et al., 2022) and (Keller Tesser and Borin, 2023).

---

[1] Metadata Server

[2] Object Storage Server

### 7.5 Parallel computing

*7.5.1 Floating point reproducibility*

Floating-point arithmetic is not associative. Therefore, when dealing with out-of-order operations, reproducibility can be lost. One method to address this problem is to implement algorithmic solutions directly to manage floating-point computations. Demmel and Nguyen contributed significantly to the topic. In 2013, (Demmel and Nguyen, 2013b) addressed the challenge of achieving reproducibility in floating-point summations, particularly in the face of dynamic scheduling and the non-associativity of floating-point operations in parallel computing environments. They proposed a technique for floating-point summation that is reproducible independent of the order of summation using the Rump algorithm (Rump et al., 2010), which is more efficient than using high-precision arithmetic. This solution trades off between efficiency and accuracy. The performance of this solution is improved with the OneReduction algorithm, from Demmel and Nguyen (Demmel and Nguyen, 2014) by relying on indexed floating-point numbers and requiring a single reduction operation to reduce the communication cost on distributed memory parallel platforms. However, these solutions did not improve accuracy. The computed results, even if reproducible, are still subject to accuracy problems. This is particularly true when addressing ill-conditioned problems (Chohra et al., 2016). Based on Stodden from (Stodden et al., 2013), the use of the double-double type or Kahan's summation can increase reproducibility without increasing computation time. In terms of fast and accurate compensated summation algorithms, the recent studies of (Blanchard et al., 2020) and (Lange, 2022) are noteworthy. Interval arithmetic is also considered as a solution (Revol and Théveny, 2014).

Some tools have also been developed to enhance reproducible floating-point arithmetic. The Intel MKL library presents conditional numerical reproducibility (CNR) (Rosenquist, 2012). This functionality curtails the utilization of instruction set extensions to guarantee consistent numerical outcomes across various architectures. However, this approach tends to considerably reduce the performance, particularly for newer architectures. In addition, it mandates a consistent thread count from one execution to the next to maintain consistent results. Verrou (Févotte and Lathuilière, 2016) is a tool built upon the Valgrind framework, which utilizes Monte Carlo arithmetic to monitor the accuracy of floating-point operations in numerical simulations without requiring source code instrumentation or recompilation. Designed for both small-scale applications and complex industrial codes, Verrou helps to diagnose inaccuracies stemming from floating-point computations, helping the verification and validation processes in industries like Electricité De France (EDF), which relies on numerical simulations for the safety and efficiency of their nuclear plants.

However, to conduct reproducible research in high-performance computing, higher-level solutions that directly work on floating-point arithmetic are preferred. Good floating-point arithmetic algorithms cannot solve all problems induced by the complex parallelization of stochastic simulations.

*7.5.2 Record and replay*

When dealing with a nondeterministic workflow such as parallel computation for high-performance computing, it can be difficult to repeat our own experiment to obtain identical bitwise results, although this is mandatory for debugging. Record and replay tools can be useful for repeating or reproducing nondeterministic computations as suggested by (Chapp et al., 2018). We synthesized the latter work to observe how such tools were initially proposed to debug large parallel computations on parallel clusters. Out-of-order execution with high-performance libraries such as the message passing interface (MPI) can indeed make computation nondeterministic. The MPI is a widely used standard for managing coarse-grained concurrencies on distributed computers.

First, certain record and replay tools have been designed for a shared memory architecture, where all processors work on the same memory. The output-deterministic replay (ODR) tool (Altekar and Stoica, 2009) is a software-only tool designed for multiprocessor programs that achieves Output-Deterministic Replay by recording only a subset of the execution data and employing a search strategy during the replay to converge to an execution that reproduces the original program outputs. By eschewing the need for a high-fidelity replica of the entire execution and sidestepping data race issues, ODR can reproduce behaviors in real multiprocessor applications such as Apache and MySQL, with an average recording overhead of 1.6 times. PRES (Park et al., 2009), as introduced by Park, uses the "record and replay" approach for debugging multiprocessors by developing "execution sketches" during recording. They iteratively explored the potential execution paths that aligned with these sketches to closely reproduce a buggy run. Most bugs were exactly replicated within the 10 replay trials. Lee's Respec (Lee et al., 2010) is a deterministic replay system that concurrently executes the replay alongside the monitored run, periodically verifying for discrepancies between the two and maintaining checkpoints of mutually accepted states. By employing speculative logging and externally deterministic replay techniques, Respec manages overhead and ensures correctness. The evaluations revealed an overhead of 18% for two-threaded programs and 55% for four-threaded applications when tested on the PARSEC and SPLASH-2 benchmark suites. ScalaMemTrace (Budanur et al., 2011) utilizes extended power regular section descriptors (EPSRDs) to deliver a compressed memory trace representation, identifying recurring behavioral patterns across the memory hierarchy. When evaluated against HPC-centric workloads, ScalaMemTrace maintained a nearly constant trace size for up to 64 threads. The replay fidelity of this tool was recognized as limited, achieving 91% accuracy on the AMG benchmark. Intel's PinPlay developed pin dynamic instrumentation framework in (Patil et al., 2010), and offers a suite of adaptable record and replay capabilities, such as subgroup replay, and ensures compatibility with other pin-affiliated tools, aiming for versatility. When assessed on HPC workloads, PinPlay registered an execution time overhead that ranged from 36 to 147 times during the recording phase. Light (Liu et al., 2015) employed a software-based method that uses satisfiability modulo theories (SMT) solvers to succinctly determine and log essential trace data for an accurate replay, focusing specifically on the flow dependence of shared memory accesses. Evaluations on a diverse set of benchmarks indicate that Light has 44% logging overhead and 10% space overhead compared to traditional techniques, making it efficient. Porridge in (Utterback et al., 2017) targets Cilk Plus programs (an extension of the C and C++ languages to support data and task parallelism). It uses a processor-oblivious "record and replay" mechanism that centers around per-lock records rather than per-thread, influencing the Cilk scheduler to align with recorded access orders through enriched direct acyclic graph representation. Evaluations spanning diverse benchmarks revealed that the recording overhead fluctuated from levels to 3.39 times, averaging at 1.62 times. Rerun (Hower and Hill, 2008) introduced a deterministic replay mechanism for multiprocessors that capitalizes on atomic episodes (sequences of instructions executed by a thread without conflicts with others) instead of recording individual memory conflicts. It requires a relatively small amount of memory per core. QuickRec (Pokam et al., 2013) is an extension of the Intel Architecture, offering hardware-assisted records and replay capabilities for multithreaded programs in multicore systems. (Chitlur et al., 2012) utilizes the QuickIA emulation platform and a modified Linux kernel. QuickRec has minimal memory log generation and performance overhead from its recording hardware, though its software stack introduces an average overhead of approximately 13%. Samsara (Ren et al., 2015) capitalized on hardware-assisted virtualization (HAV) extensions using an approach for deterministic replay in multiprocessor systems without the need for hardware alterations. By employing a chunk-based recording scheme that avoids all memory access detections (a primary source of overhead in preceding methods), they argued that Samsara markedly diminishes the log file size to 1/70th and reduces the recording overhead from approximately 10 times to an average of 2.3 times when compared to conventional software-only solutions. More recently, in 2017, Castor (Mashtizadeh et al., 2017) offered a default-on record and replay solution for multi-core applications, consistently emphasizing minimal and predictable overheads. Notably, whereas Castor achieves a low recording overhead for the majority of the PARSEC benchmarks, the Radiosity benchmark sees overheads reaching 25% on a 10-thread run owing to cache impacts and log aggregation challenges. Castor can work in the C, C++, and Go languages.

Other tools are designed for distributed memory architectures, where each processor has its own private memory, and processors need to communicate by passing messages between each other, most often using the MPI library. Scala-H-Trace (Wu et al., 2011) employs aggressive trace compression to capture variations in communication and I/O parameters via probabilistic histograms, ensuring near-constant trace file sizes even in variable patterns. Even with its compressed trace approach, Scala-H-Trace deterministically replays these probabilistic traces without deadlocks, closely resembling the original applications, with replay times within 12–15% of the original runtimes. Similar to Scala-H-Trace, Scala-Trace II (Wu and Mueller, 2013) uses trace compression techniques that are particularly effective for HPC applications with irregular behaviors and employs a low-level encoding scheme that enhances data elasticity and interpretability. It achieves substantial trace compression even in applications with inconsistent time-step behavior, and in evaluations, demonstrates an average execution time error of 5.7% in its probabilistic replay when compared to the original executions. (Guermouche et al., 2011) introduced an uncoordinated checkpointing protocol tailored for send-deterministic MPI HPC applications that log only selected messages and do not mandate a full process restart post-failure. (Meneses et al., 2010) used a technique designed to mitigate the memory overhead inherent in in-memory message logging by clustering processors into teams, where only inter-team messages require logging. Xue et al. (Xue et al., 2009) introduced MPIWiz based on the subgroup reproducible replay (SRR) method, a hybrid deterministic replay technique for MPI applications that balances the advantages and limitations of both data and order replays. Using MPIWiz, the system captures message content between process groups and only message orderings within a group, showing a 27% increase in execution time during recording, and allowing for a replay of merely 53% of the application's base execution time. Gioachin et al. (Gioachin et al., 2010) presented a three-step hybrid replay mechanism in which the initial passes adopt a minimalistic order replay, subsequently transitioning to more intensive data replays but confined to progressively fewer processes, facilitating targeted bug tracing. Rex (Perianayagam et al., 2010), introduced by Perianayagam et al., is a toolset designed to comprehensively record, archive, and replay software experiments, ensuring the fidelity of reproduction despite potential changes in external data sets, unavailable original software, or undocumented input parameters. This adds a minimal execution time overhead of approximately 1.6% and an archiving space overhead ranging from 5 to 7 GB. The complete state-of-the-art records and replay tools can be found in (Chapp et al., 2018).

Numerous tools have recently been proposed because the nondeterminism problem is ubiquitous in high-performance computing. Many of these tools have a non-negligible negative effect on the performance; therefore, they are quite compromised. However, it is the last possible solution tool left for repeatability and reproducibility, which are mandatory for debugging; many of these tools were developed for debugging purposes, not for reproducibility. Nevertheless, in high performance computing, this is an existing solution that enables reproducible experimentation.

## 7.6 How do we deal with silent or 'soft errors'

Silent errors are ubiquitous in high-performance computing systems, particularly on the exascale. This phenomenon has been studied, and several solutions have been proposed.

At the memory level, error correction code (ECC) technology is designed to protect RAM memory. ECC memory is a specialized form of computer data storage that employs an error correction code to identify and rectify instances of n-bit data corruption within the memory. It has applications in situations where data corruption is unacceptable, such as in industrial control setups, critical databases, and essential memory caches. ECC memory is designed to prevent single-bit errors from affecting memory integrity, ensuring that the read data match the originally written data, even if a bit has been inadvertently altered. By contrast, most regular non-ECC memories lack error detection and correction capabilities, with only certain non-ECC memory configurations featuring parity support, which enables error detection without correction. In (Baumann, 2005), Baumann work on the error-rate observations for non-ECC protected SRAM. "*This is calculated using a Soft Error Rate (SER) from radiation resulting in an estimated 50000 FIT (Failure-In-Time: One FIT is equivalent to one failure in 1 billion device hours). Hence, they recommend using ECC which reduces the error rate by 1000x for SRAMs.*" (Dixit et al., 2021). Other hardware countermeasures against silent errors include redundancy and parity. Redundancy involves duplicating critical components or data such that if an error occurs, the redundant component or data can be seamlessly removed. Parity, which is a simpler form of error checking, adds an extra bit to a set of data to ensure an even or odd parity (the sum of the bits being even or odd). If an error occurs, the parity bit changes, which indicates a problem.

Software solutions also exist. (Fiala et al., 2012) proposed an MPI library to handle silent errors, based on redundancy. (Benson et al., 2015) proposed handling of silent errors using a new paradigm for detecting such errors at the application level. This approach centers on frequently comparing computed values with those generated by low-cost checking computations. Error detectors were constructed by examining the differences between two sets of output sequences. The authors used numerical analysis to identify suitable checking computations for solving the initial value problems in ordinary differential equations (ODEs) and partial differential equations (PDEs). Specifically, the solution employs methods such as the Runge-Kutta and linear multistep methods for ODEs and implicit and explicit finite-difference schemes for PDEs. The authors used examples such as the heat equation and Navier-Stokes equations to demonstrate their approach. Importantly, through tests involving deliberately introduced errors, the proposed approach effectively identified nearly all meaningful errors without causing a substantial decrease in performance speed. (Hoemmen and Heroux, 2011) proposed a resilient version of the GMRES iterative method. This resilient version is designed to correct errors and enhance the error tolerance of the method. (Bronevetsky and de Supinski, 2008) and (Casas et al., 2012) suggested that empirical studies have shown that some iterative methods might be vulnerable to errors. These studies indicate that not all iterative methods inherently correct errors. (Huang and Abraham, 1984) introduced checksum methods to enhance the integrity of matrix multiplication and to effectively detect errors during the process. (Du et al., 2012) proposed the application of checksum methods to improve the robustness of high-performance LU (lower-upper) decomposition, thereby detecting potential errors in the LU factorization process. In (Aupy et al., 2013), a solution to handle silent data corruption errors that revolves around revisiting traditional checkpointing and rollback recovery strategies was proposed. The focus is on addressing latent errors that are not immediately detected. They introduced two models to handle these errors. In the first model, errors are detected after a delay, following a probability distribution that is often represented as an exponential distribution. The solution computes the optimal period for checkpointing, aiming to minimize the waste of time when the nodes are not engaged in useful computations. Because only a limited number of checkpoints can be stored in memory, there is a possibility of irrecoverable failure owing to limited resources. In these cases, the minimum period required to achieve an acceptable level of risk is determined. In the second model, the errors were detected using a verification mechanism. Unlike in the first model, there is no risk of irrecoverable failure because the verification mechanism ensures that errors are detected. However, the overhead introduced by this verification mechanism was additional computations.

The primary goal is to find an optimal checkpointing period that minimizes time consumption, considering the trade-off between early error detection, limited checkpoint storage, and the overhead introduced by verification mechanisms. These models were applied to real-world scenarios, and various applications and architectural parameters to demonstrate their feasibility and effectiveness were considered. More recent technologies, such as machine learning, may offer new solutions. (Wang et al., 2018) proposed a neural network detector that could detect silent data corruption even after multiple iterations after the data were injected. The effectiveness of the proposed neural network detector was evaluated using six flash applications and two Mantevo mini-applications. The experimental results demonstrated that this detector can successfully identify more than 89% of silent data corruptions, while maintaining a low false-positive rate of less than 2%. More information on this subject can be found in a survey of techniques for modeling and improving the reliability of computing systems (Mittal and Vetter, 2015).

# 8 UNRESOLVED PROBLEMS IN REPRODUCIBLE RESEARCH FOR HPC

## 8.1 Portability of algorithms and particularly PRNGs

Verifying the correctness of algorithms in high performance computing (HPC) is a challenging task. A prime example is the inconsistency observed when a pseudo-random number generator (PRNG) is implemented using different programming languages or technologies. Among the top PRNGs, Philox, discussed previously, is an example where you have to be careful, and MLFG is another example in which we encountered portability issues. When a PRNG is initialized with the same seed (or initial status), we expect an identical stream of numbers across various computing environments and libraries (e.g., Numpy and TensorFlow). However, variations arise, casting doubt on the portability and consistency of the algorithms. This raises the question of the reliability of the scientific results derived from such algorithms. The foundation of deterministic computational science hinges on the predictability and reliability of algorithms. The observed disparities in PRNG outputs underscore a broader issue: without standardized implementation and rigorous cross-platform verification, how can we ensure the integrity of our algorithmic results? This calls for a concerted effort to develop and adhere to the standardization of PRNG implementation. This necessitates the creation of robust verification frameworks that can ensure algorithmic fidelity and portability across diverse computational environments, which are crucial for the advancement of science.

## 8.2 Data Reproducibility in Big Data Infrastructures

The proliferation of big data in HPC has introduced multifaceted challenges in data reproducibility. As datasets grow, conventional data versioning systems struggle to cope, leading to significant hurdles in maintaining a consistent and reproducible data environment. This is not just a technical issue, but also a methodological one, where the need for scalable and efficient management of these voluminous datasets is paramount. Furthermore, with the increasing emphasis on data-driven science, it is imperative to establish robust data-sharing practices tailored to the immense scale and complexity of HPC big data ecosystems. These practices must not only facilitate sharing, but also ensure that data integrity and reproducibility are preserved. Updating large databases can lead to a loss of reproducibility, and how do we deal with the archiving of versions? Addressing these challenges is essential to enable collaborative scientific discovery and maintain the credibility of computational research.

### 8.3 Performance Reproducibility and Optimization

The quest for high-speed computation in HPC often leads to a tradeoff between reproducibility, particularly when dealing with parallel and distributed systems. Optimization techniques that boost performance, such as tailoring code to specific architectures or leveraging concurrent execution, can render the results less reproducible. This dichotomy presents a critical open problem: how can we balance the pursuit of peak performance with the assurance of consistent results? It is a delicate balancing act that requires a deeper understanding of the interplay between system architecture, optimization strategies, and the nature of computational tasks. An example given in the previous section is, for instance, the use of fused multiply-add (FMA) or advanced vector extensions (AVX), which can lead to a loss in reproducibility for HPC. This assumes, for instance, that clusters and supercomputers disable AVX and/or FMA instructions. The latter corresponds to a multiply-add floating-point operation performed in a single step and with a single rounding; the change in the order of floating-point operations and precision leads to a lack of repeatability. Research in this domain is vital for developing new methods that can guarantee reproducible outcomes without sacrificing the performance of the HPC systems.

### 8.4 Education, Community Collaboration, and Workflow Integration

The integration of reproducibility into the fabric of scientific inquiry is a challenge that transcends technical solutions and affects educational, collaborative, and procedural aspects. There is a pressing need to inculcate the principles of reproducible research within educational curriculums to cultivate a new generation of scientists who naturally integrate these practices into their work. Beyond academia, fostering community standards and collaboration is essential for developing and maintaining reproducible workflows. This involves not only the individual researcher but the entire scientific ecosystem, including publishers, funders, and institutions. The goal is to create a scientific culture in which reproducibility is not an afterthought, but a fundamental component of the scientific process, from data acquisition to analysis.

### 8.5 Addressing Reproducibility in Emerging Computational Models

Emerging computational models, particularly quantum computing, present unprecedented challenges in terms of reproducibility. For instance, quantum computing operates under a set of principles different from those of classical computing, leading to new types of errors and uncertainties. This field is still in its infancy and several technical options are being explored to create reliable qubits for quantum circuits. Thousands of qubits are required to model a single perfect qubit, and experiments that perform properly on quantum simulators are often unreproducible on real quantum machines (Hill et al., 2023). These challenges are compounded by the probabilistic nature of quantum computing, which renders repeatability and reproducibility even more challenging. Current tools and practices are ill-equipped to address these novel issues, necessitating concerted research efforts to understand and overcome the barriers to reproducibility in this revolutionary computational paradigm.

### CONCLUSION

Achieving reproducibility in high-performance computing is challenging. Tools often prioritize user-friendliness over performance optimization, as observed with the FAIR principles designed for disciplines such as biology. This represents a tradeoff between usability and performance. Therefore, it is essential to differentiate between general reproducible research and reproducibility within an HPC context.

In the vanguard of scientific exploration, HPC embodies the cutting edge yet struggles with the fundamental principle of reproducibility. This study embarked on a journey through the intricate landscape of HPC, scrutinizing the reproducibility crisis that plagues numerous scientific domains, and not just this field. We have emphasized the pivotal role of software engineering, versioning, workflow management, and scientific culture in influencing reproducibility and proposed robust solutions such as literate programming, advanced workflow management, and containerization technologies such as Guix and Apptainer.

In our opinion, the last remaining problem for reproducible research is the scientific culture of computer scientists. Funding agencies should focus on emphasizing and incentivizing reproducible research and encouraging researchers to leverage existing tools. With reasonable efforts, it could become standard practice to accompany scientific publications with related artifacts. Because computer science is a relatively young research field, many practitioners lack epistemological training. This is often observed when computer scientists use the term methodology instead of method. However, with continued advocacy for reproducible research, the well-established practice of maintaining laboratory notebooks in biology may find its counterpart in computer science, the computational notebooks.

In the case of reproducibility in high performance computing, it is more complicated. Exascale computing is now a reality. Machines are becoming much denser, and even if they show the best performance, they have limited uptime, which is often due to silent errors. Increased computing power is consistently required to conduct extensive explorations using complex programs. The complete design of computer experiments with all the factors and levels (a combination of input parameters) is often intractable. Therefore, increasing the sampling of all possible experiments is desirable and requires intensive computing. Even if the main goal is to enhance speed, it necessitates the deployment of extensive parallel architectures and optimizations, and we discovered in the last decade that this is too often achieved at the expense of repeatability. High-cost computations are those that we do not want to perform many times (Hinsen, 2021). However, we must trust the obtained results. Achieving repeatability in HPC often results in a loss of performance when the user accepts to disable certain optimizations that appear unreliable. We believe that the remaining issues regarding reproducibility are related to large optimized parallel simulations. For simpler computation, tools such as Guix or Apptainer (Singularity) have been shown to be efficient, with near-native performance. Currently, regardless of whether a shared memory or distributed memory platform is used, if large parallel computations are heavily optimized, a lack of numerical repeatability is experienced.

Emerging challenges also stem from the AI and big data domains. A significant fraction of the global computational resources is currently dedicated to training AI models and storing vast datasets. Reproducibility not only in terms of numerical results, but also in terms of performance, might be challenging; however, it is essential to maintain explainability. In addition, we must address the imperative to decrease energy consumption, and each new system comes with a solution in this respect.

Finally, a new form of high-performance computing is available through quantum computing. This technology is perhaps the one with the most open and challenging problems. At our disposal, we do not have a general quantum computer with perfect qubits; numerous technical options have been tested to provide excellent qubits, but they are not perfect; thus, it is currently difficult to design reliable circuits. If quantum simulators work properly, the use of hybrid quantum machines encounter several challenges in achieving statistical reproducibility.

In conclusion, computers serve as fundamental tools not only for computer scientists, but also for a multitude of scientific disciplines. Similar to the precise and systematic approach to metrology applied to the instruments used by biologists and physicists, scientists must acknowledge the inherent imperfections and uncertainties of computational tools.

It is imperative that the principles of scientific methodology, including the tenets of reproducibility, be integrated into the curriculum of computer science education programs.

**CRediT authorship contribution statement**

**Benjamin Antunes**: Writing–original draft, Conceptualization, Investigation.

**David R.C. Hill**: Supervision, Writing – review & editing, Validation.

**Declaration of competing interest**

The authors declare that they have no known competing financial interests or personal relationships that could have appeared to influence the work reported in this paper.

**Acknowledgement**

We would like to thank Michael R. Crusoe for his interesting suggestions about workflows.

Antunes, B., & Hill, D. (2023). Identifying quality mersenne twister streams for parallel stochastic simulations. *2023 Winter Simulation Conference (WSC)*.

Aupy, G., Benoit, A., Hérault, T., Robert, Y., Vivien, F., & Zaidouni, D. (2013). On the combination of silent error detection and checkpointing. *2013 IEEE 19th Pacific Rim International Symposium on Dependable Computing*, 11-20.

Baier, T., & Neuwirth, E. (2007). Excel : Com. *Computational statistics*, *22*(1), 91-108.

Bajpai, V., Bonaventure, O., Claffy, K., & Karrenberg, D. (2019). Encouraging reproducibility in scientific research of the internet. *Dagstuhl reports*, *8*(10).

Bajpai, V., Brunstrom, A., Feldmann, A., Kellerer, W., Pras, A., Schulzrinne, H., Smaragdakis, G., Wählisch, M., & Wehrle, K. (2019). The Dagstuhl beginners guide to reproducibility for experimental networking research. *ACM SIGCOMM Computer Communication Review*, *49*(1), 24-30.

Bajpai, V., Kühlewind, M., Ott, J., Schönwälder, J., Sperotto, A., & Trammell, B. (2017). Challenges with reproducibility. *Proceedings of the Reproducibility Workshop*, 1-4.

Baker, M. (2016). Reproducibility crisis. *Nature*, *533*(26), 353-366.

Barba, L. A. (2018). Terminologies for reproducible research. *arXiv preprint arXiv:1802.03311*.

Baumann, R. C. (2005). Radiation-induced soft errors in advanced semiconductor technologies. *IEEE Transactions on Device and materials reliability*, *5*(3), 305-316.

Baumer, B., Cetinkaya-Rundel, M., Bray, A., Loi, L., & Horton, N. J. (2014). R Markdown : Integrating a reproducible analysis tool into introductory statistics. *arXiv preprint arXiv:1402.1894*.

Begley, C. G., & Ellis, L. M. (2012). Raise standards for preclinical cancer research. *Nature*, *483*(7391), 531-533.

Ben-David, S., Hrubeš, P., Moran, S., Shpilka, A., & Yehudayoff, A. (2019). Learnability can be undecidable. *Nature Machine Intelligence*, *1*(1), 44-48.

Benedicic, L., Cruz, F. A., Madonna, A., & Mariotti, K. (2019). Sarus : Highly scalable docker containers for hpc systems. *High Performance Computing: ISC High Performance 2019 International Workshops, Frankfurt, Germany, June 16-20, 2019, Revised Selected Papers 34*, 46-60.

Benson, A. R., Schmit, S., & Schreiber, R. (2015). Silent error detection in numerical time-stepping schemes. *The International Journal of High Performance Computing Applications*, *29*(4), 403-421.

Beserra, D., Oliveira, F., Araujo, J., Fernandes, F., Araújo, A., Endo, P., Maciel, P., & Moreno, E. D. (2015). Performance evaluation of hypervisors for hpc applications. *2015 IEEE International Conference on Systems, Man, and Cybernetics*, 846-851.

Blanchard, P., Higham, N. J., & Mary, T. (2020). A class of fast and accurate summation algorithms. *SIAM journal on scientific computing*, *42*(3), A1541-A1557.

Boettiger, C. (2015). An introduction to Docker for reproducible research. *ACM SIGOPS Operating Systems Review*, *49*(1), 71-79.

Booch, G., Jacobson, I., & Rumbaugh, J. (1996). The unified modeling language. *Unix Review*, *14*(13), 5.

Borthakur, D. (2007). The hadoop distributed file system : Architecture and design. *Hadoop Project Website*, *11*(2007), 21.

Boyer, A. F. (2022). *Contributions to Computing needs in High Energy Physics Offline Activities : Towards an efficient exploitation of heterogeneous, distributed and shared Computing Resources*. CERN.
40

## AUTHOR BIOGRAPHIES

**BENJAMIN ANTUNES** is a Phd Student at Clermont Auvergne University (UCA). He holds a Master in Computer Science (head of the list). His thesis subject is about the reproducibility of numerical results in the context of high performance computing. He is espacially working on reproducibility issues in parallel stochastic computing. His email address is benjamin.antunes@uca.fr and his homepage is https://perso.isima.fr/~beantunes/

**DAVID HILL** is doing his research at the French Centre for National Research (CNRS) in the LIMOS laboratory (UMR 6158). He earned his Ph.D. in 1993 and Research Director Habilitation in 2000 both from Blaise Pascal University and later became Vice President of this University (2008-2012). He is also past director of a French Regional Computing Center (CRRI) (2008-2010) and was appointed two times deputy director of the ISIMA Engineering Institute of Computer Science – part of Clermont Auvergne INP, #1 Technology Hub in Central France (2005-2007 ; 2018-2021). He is now Director of an international graduate track at Clermont Auvergne INP. Prof Hill has authored or co-authored more than 250 papers and has also published several scientific books. He recently supervised research at CERN in High Performance Computing. ( https://isima.fr/~hill/  -  david.hill@uca.fr )